Preprinted manuscript

# Estimation of symbiotic bacterial structure in a sustainable seagrass ecosystem on recycled management

Hirokuni Miyamoto*[1,2,3,4], Nobuhiro Kawachi[5], Atsushi Kurotani[6], Shigeharu Moriya[7], Wataru Suda[2], Kenta Suzuki[8], Makiko Matsuura[1,4], Naoko Tsuji[4], Teruno Nakaguma[1,3,4], Chitose Ishii[2,4], Arisa Tsuboi[3], Chie Shindo[6], Tamotsu Kato[2], Motoaki Udagawa[9], Takashi Satoh[10], Satoshi Wada[7], Hiroshi Masuya[8], Hisashi Miyamoto[4,11], Hiroshi Ohno[2], Jun Kikuchi*[12]

**Affiliations:**
[1] Graduate School of Horticulture, Chiba University: Matsudo, Chiba 271-8501, Japan
[2] RIKEN Center for Integrated Medical Science: Yokohama, Kanagawa, 230-0045, Japan
[3] Japan Eco-science (Nikkan Kagaku) Co. Ltd.: Chiba, Chiba 263-8522, Japan
[4] Sermas Co., Ltd.: Ichikawa, Chiba 272-0033, Japan
[5] Kawachi Suisan Ltd.: Saiki, Oita, 876-2302, Japan
[6] Research Center for Agricultural Information Technology, National Agriculture and Food Research Organization, Tsukuba, Ibaraki, Japan, 305-0856
[7] RIKEN, Center for Advanced Photonics, Wako, Saitama, Japan, 351-0198
[8] RIKEN, BioResource Research Center, Tsukuba, Ibaraki, Japan, 305-0074
[9] Keiyo Gas Energy Solution Co. Ltd.: Ichikawa, Chiba 272-0033, Japan
[10] Division of Hematology, Kitasato University School of Allied Health Sciences, Sagamihara, Kanagawa 252-0329, Japan
[11] Miroku Co.Ltd.: Kitsuki, Oita 873-0021, Japan
[12] RIKEN Center for Sustainable Resource Science: Yokohama, Kanagawa, 230-0045, Japan

* Co-corresponding authors
Correspondence: Hirokuni Miyamoto Ph.D., Chiba University, RIKEN, Sermas Co., Ltd., and Japan Eco-science Co., Ltd.
Tel: +81-43-290-3947, Fax: +81-43-290-3947
E-mail: hirokuni.miyamoto@riken.jp, h-miyamoto@faculty.chiba-u.jp

Correspondence: Jun Kikuchi Ph.D., RIKEN
Tel: +81-43-290-3942, Fax: +81-43-290-3942
E-mail: jun.kikuchi@riken.jp

**Abstract**
Seagrass meadows play an essential role in blue carbon and aquatic ecosystem services. However, methods for the flourishing of seagrass are still being explored. Here, data from 49 public coastal surveys on the distribution of seagrass and seaweed around the onshore aquaculture facilities are revalidated, and an exceptional area where the seagrass *Zostera marina* thrives was found. The bacterial population there showed an apparent decrease in the pathogen candidates belonging to the order Flavobacteriales. Moreover, structure equation modeling and a linear non-Gaussian acyclic model based on the machine learning data estimated an optimal symbiotic bacterial group candidate for seagrass growth as follows: the Lachnospiraceae and Ruminococcaceae families as gut-inhabitant bacteria, Rhodobacteraceae as photosynthetic bacteria, and Desulfobulbaceae as cable bacteria modulating oxygen or nitrate reduction and oxidation of sulfide. These observations confer a novel perspective on the seagrass symbiotic bacterial structures critical for blue carbon and low-pathogenic marine ecosystems in aquaculture.

## Introduction

Sustainable food production, related to essential improvements, is an urgent worldwide issue [1]. The global food crisis is a critical issue, and the fish industry is expected to grow as food production grows [2] because terrestrial food security is likely to hurt climate change and the entire ecosystem due to global land use [3]. At the moment, seafood food is less than 20% of total meat. Unsustainable practices, regulatory barriers, distorted incentives, and other restrictions limit seafood production. However, policy reforms and technological innovations are required to protect the marine environment [4].

Seagrass plays an essential role in aquatic ecosystem services involved in fish growth of fish [5] and environmental conservation. The potential of seagrass to suppress greenhouse gases is approximately 40 times more efficient in capturing organic carbon than land forest soils [6,7], called blue carbon [8]. It is becoming clear that the adverse effects of carbon dioxide are even greater than previously assumed [9], and the presence of seagrass should become increasingly important. The role of seagrass is closely related to the concept of planetary boundaries [10], the framework of which defines the environmental limits that ensure the safety of humanity and the stability of the Earth system. Therefore, improving the nitrogen and phosphate cycle and biodiversity are considered urgent issues.

Recent research has pointed out the possibility of efficient nitrogen circulation by seagrass [11] and sediment [12], and phosphorus [13]. Significantly, seagrass affects the reduction of pathogens [14]. Although seaweed overgrowth is essential, as mentioned above, seagrass death is frequently observed [15]. Since the 20th century, carbon dioxide emissions have increased due to seagrass loss [16]. Furthermore, ocean acidification is progressing worldwide and is considered a problem preserving seagrass ecosystems [17]. It has also been pointed out that global warming also suppresses the overgrowth of seagrass, making it difficult to transplant itself[18]. However, seagrass pastures can locally alleviate low pH conditions for extended periods, with essential implications for the conservation and management of coastal ecosystems. Therefore, efficient methods for the cultivation of seagrass have been explored worldwide.

In the current study, 49 coasts around Saiki city, Oita Prefecture, a particular area with abundant onshore aquaculture facilities, were validated to understand the distribution and characteristics of seagrass sediment using machine learning. A characteristic of the seagrass growing area was the absence of onshore aquaculture facilities. As the only exception, seagrass overgrowth was found in an onshore aquaculture facility. The physicochemical properties of the seagrass sediment around the coast downstream of the aquaculture facility and the characteristics of the symbiotic bacterial population of the seagrass sediment were evaluated in detail. Generally, sea sediment analysis can clarify the deterioration in estuaries[19,20] and marine ecosystems[21-24]. Therefore, these sediment data were evaluated using a procedure combined with machine learning, structure equation modeling (SEM), and statistical causal inference (Fig. S1). SEM and statistical causal analyses based on machine learning data predicted the symbiotic bacterial groups that could be strongly involved in the seagrass growth around aquaculture facility. These observations provide critical points of view for the development of sustainable technologies to preserve marine ecosystems and the global environment.

## Materials and Methods

### Reinvestigation based on historical data

The location of seaweeds and seagrass along the coasts (E131°94′-E132°00′, N36°87′-N32°81′) around Saiki city, Oita Prefecture, was publicly surveyed in 2004 [25] and 2016 [26]. Data rearrangement was performed for the following association analysis, an elementary unsupervised machine learning method. The map of Japan in Figs. 1a and S1 was visualized using the packages 'tidyverse', 'maps', 'jpndistrict' and 'countrycode' in the R software (version 4.0.5) (https://cran.r-project.org). The other maps in Figs. 1a, S1, and S2 were visualized using the software package 'new pec' (https://www.jha.or.jp/en/jha/).

### Association analyses

Association analysis is suitable for different classifications in different layers and is a well-established method used in market research for market basket analysis [27-30]. It can be applied to conditions that are otherwise difficult to compare horizontally to obtain and classify relevant components. For example, to predict components associated with onshore aquaculture facilities or seagrass, association analysis was performed as previously reported for other subjects [27-30]. In summary, association analysis is a rudimentary method of inferring a "target," which is an effect, from a "source," which is a cause. In this case, "cause" and "target" are represented by x and y, respectively, and the association analysis equation is defined by probability (P) as follows support $(x \Rightarrow y) = P(x \cap y)$.

The "support" is denoted as $P(xy)$ and the probability (P) that x and y co-occur.

Confidence $(x \Rightarrow y) = P(x \cap y) / P(x)$.

The "confidence" is also denoted as $P(xy)/P(x)$ and is the conditional probability (P) that y will occur after x occurs.

lift $(x \Rightarrow y) = P(x \cap y)/P(x)P(y)$.

The "lift" is $P(xy)/P(x)P(y)$, a measure of association/independence.

A value of lift > 1 indicates a positive association, and a value of lift < 1 indicates a negative association. Here, the association rule was determined using the criterion values of support, confidence, and lift (support = 0.063, confidence = 0.25, maxlen = 2, lift > 1.5). All target-based information was used, i.e., presence/absence of seaweed/seagrass, its location, bacterial classification data, and joint data on the presence/absence of seaweeds. All data were calculated based on the median (M) of each data category and sorted by binarization such as 0 (< M) or 1 (> M). The dataset was ruled by the R software package "arules". The association network was drawn by Force Atlas with Noverlap in Gephi 0.9.2 (https://gephi.org).

### Sample preparation

Sediment and seawater near areas selected based on previous surveys [25,26] were randomly collected in 2019-2021. For example, the bottom sediment inhabited by the *Zostera marina of the seagrass* was collected at the location shown in Fig. S6. The collected samples were frozen at -60 °C until the experiment was performed.

### Identification of the seagrass Zostera marina

The DNA sequence of the seagrass was identified by the experimental protocol of Reso Co., Ltd., Japan. The procedure was performed as follows. DNA was extracted from plant pieces using DNA rinse-E (https://rizo.co.jp/image/96.pdf) and PCR with ITS1-18S_forward, CGTAACAAGGTTTCCGTAGG, and ITS1-5.8S reverse, CGTTCTTCATCGATGCGAGA, as a *Zostera* primer set (ITS1-5.8S, ITS1-18S). The nucleotide sequence of the amplicons was determined using an Applied Biosystems 3730xl DNA analyzer (Macrogen Japan Co., Ltd.). The determined nucleotide sequence was collated with the DNA database to identify the species. Sequencing data can be found under accession number LC636204 in DDBJ with the entry ID of 60b877bf3a01a50012fc72e0.entry1.

### Physicochemical analyses

Physicochemical components were determined by inductively coupled plasma (ICP) equipment and a Soil & Plant Clinical Analyzer SFP-4j (Fujdaira, Co., Ltd.), owned by Agrokanesho Co., Ltd., Japan. These determinations were conducted on the basis of the manufacturer's experimental protocol. The physicochemical indices were visualized using Prism software (version 9.2.0).

### Stable isotope analysis of carbon and nitrogen

Carbon and nitrogen isotopic compositions were determined using the manufacturer's protocols [31-34] with some modifications by DELTA V Advantage (Thermo Fisher Scientific, USA) and Flash EA1112 (Thermo Fisher Scientific, USA). These procedures were conducted by Shoko

Science Co., Ltd., Japan. In summary, the total carbon and total nitrogen of the samples were measured using the Flash EA1112 as follows: The oxidation and reduction reactors were heated to 1,000 °C and 680 °C, respectively. The flow rate of the carrier gas (He) was approximately 100 mL/min. The length of the separation column was 3 m, and the oven temperature of the column was set at 35 °C. Acetanilide (Kishida Chemical Co., Ltd., Japan) was used as a standard. Stable carbon and nitrogen isotopes were measured using an elemental analyzer/isotope ratio mass spectrometer (EA/IRMS), a unit of DELTA V Advantage interfaced with FlashEA 1112, as follows: The oxidation and reduction reactors were heated to 1,000 °C and 680 °C, respectively. The flow rate of the carrier gas (He) was about 100 mL/min. The length of the separation column was 3 m, and the oven temperature of the column was set at 35 °C. $CO_2$ and $N_2$ were used as reference gases to detect stable carbon and nitrogen isotopes, respectively. Alanine (19.6‰), histidine (10.7‰), and glycine (33.8‰) were used as working standards for carbon ($\delta^{13}C$), Vienna Pee Dee Belemnite (VPDB). Alanine (1.58‰), alanine (9.97‰), and glycine (20.6‰) were used for the working standard stable nitrogen air ($\delta^{15}N$).

*Meta-sequence analysis of the 16S rRNA gene of bacteria*
Each sediment sample was aliquoted into a 15 ml tube. 5 ml of sterile water was added, the sample was well suspended by vortexing, filtered through filter paper (Advantech 5C) (Advantech Co., Ltd., USA), and the filtrate was collected in a new 15 ml tube. Power Soil Pro Kit (QIAGEN Co., Ltd., USA) was used to extract DNA according to the manufacturer's protocol. The DNA sample was amplicon generated using the MetaVx TM Library Preparation Kit (GENEWIZ, Inc., South Plainfield, NJ, USA), and the V3-V4 region (341F-806R) of the bacterial 16S rRNA gene was amplified and sequenced with Illumina MiSeq according to the method in previous studies [35]. The obtained sequences were filtered by Trimomatic (http://www.usadellab.org/cms/?page=trimmomatic). As previously reported [36], 2,000 trimmed reads per sample were analyzed with QIIME 1.9.1. Filtered sequences were clustered into operational taxonomic units (OTUs); OTU classification was defined at the 97% similarity level. α-diversity, β-diversity, bacterial communities and heatmaps were generated using the software R package 'genefilter', 'gplots', 'ggplot2', 'RColorBrewer', 'pheatmap', 'ape', 'base', 'dplyr', 'easyGgplot2', 'knitr', 'ggthemes', 'phyloseq' and 'vegan' were used for visualization. Microsoft 365, and Prism software (version 9.2.0). The number of observed OTUs and Chao1, Shannon, and Simpson values were evaluated as α-diversity. β-Diversities were estimated using principal coordinate analysis (PCoA) for weighted or unweighted UniFrac distances, and Bray-Curtis distance based on OTU distribution among samples as previously described [37]. These diversities were analyzed by the function 'adonis' in the R software packages "vegan" and "MASS". The relative abundance of individual phylum, class, order, family and genus was compared within the major communities (comparison in > 0.1% of the total bacterial community). Sequence data can be found under the BioProject accession number PRJDB12183 in DDBJ with the temporary submission ID DRA012976.

*Linear discriminant analysis*
Linear discriminant analysis (LDA), an elementary method of supervised machine learning, supports high-dimensional class comparisons with a particular focus on meta-analyses. Here, the LDA effect size (LEfSe) was used to identify genomic taxa that characterize the differences between the experimental conditions. Galaxy constructed LDA score plots and a LEfSe-based cladogram (https://huttenhower.sph.harvard.edu/galaxy/), as described in a previous overview [38]. Briefly, the populations of members of the predominant bacterial community (>1% of the total bacterial community) were analyzed. Figures were visualized by the Kruskal-Wallis factorial test between classes and the Wilcoxon pairwise test between subclasses (set at 0.05) as nonparametric analyzes. Pairwise comparisons between subclasses to be performed only among subclasses with the same name were set to "Yes". The multiclass analysis strategy was set as 'All-against-All (more strict)". The logarithmic LDA score threshold for discriminative features was set at 4.0.

*Energy landscape analysis*
Energy landscape analysis (ELA) as a machine learning method was performed as previously described [39]. In summary, ELA can construct landscapes that grasp the stability of community compositions across environmental gradients. ELA is based on an extended pairwise maximum entropy model that explains the probability of the occurrence of the ecological state of sample $k$, $\sigma^{(k)}$ given the environmental condition $\epsilon^{(k)}$; as the ecological state, we combined the presence/absence status of selected taxa and the levels of physiological factors, $\sigma^{(k)} = (\sigma_1^{(k)}, \sigma_2^{(k)}, ... \sigma_N^{(k)}) = (\sigma_1^{(k)}, ..., \sigma_{n_m}^{(k)}, \sigma_{N-n_c}^{(k)} ... \sigma_N^{(k)})$ where $n_m$ is the number of bacterial taxa and $n_c$ is the number of chemicals, as an environmental condition. Here two environmental factors represent (1) or without (0) the distance from the targeted land-based aquaculture facility ($\epsilon_{d,i}^{(k)}$: Nos. 31 and 31_100 m (Fig S6) are converted to 1 and 0) and the seasonal stages converted to the 0-1 range ($\epsilon_{s,i}^{(k)}$: February, November, and June are converted to 0, 0.375 and 1, respectively) were combined as $\epsilon^{(k)} = (\epsilon_{d,i}^{(k)}, \epsilon_{s,i}^{(k)})$. The model can be written as

(I) $P(\sigma^{(k)}|\epsilon) = \frac{e^{-E(\sigma^{(k)}|\epsilon)}}{\Sigma e^{-E(\sigma^{(k)}|\epsilon)}}$,

(II) $E(\sigma^{(k)}|\epsilon^{(k)}) = -(\Sigma_i \Sigma_j J_{ij} \sigma_i^{(k)} \sigma_j^{(k)} + \Sigma_i g_i^{distance} \epsilon_i^{distance} \sigma_i^{(k)} + \Sigma_i g_i^{season} \epsilon_i^{season} \sigma_i^{(k)} + \Sigma_i h_i \sigma_i^{(k)})$.

In addition, $P(\sigma^{(k)}|\epsilon)$ is the probability for the occurrence of an ecological state $\sigma^{(k)}$. Eq. (I) shows that the probability is high when energy $E(\sigma^{(k)}|\epsilon)$ is low and *vice versa*. In eq. (II), $E(\sigma^{(k)}|\epsilon)$ is defined as the sum of the effect of interaction among components, antibiotic treatment, growth stages, and the net effect of unobserved environmental factors. Parameters in Eq. (II), namely, $J_{ij}$, $g_i^s$, $g_i^d$ and $h_i$, indicate the effect of the relationship among components ($J_{ij} > 0$ favors and $J_{ij} < 0$ disfavors the cooccurrence of components $i$ and $j$), the effect of the season on component $i$ (the conditions around summer (June)($g_i^s > 0$) or the conditions around winter (February) ($g_i^s < 0$) affects the occurrence of component $i$), the effect of the distance away from the targeted facility (around Area No.31 in Fig S6) on component $i$ (component $i$ favors the closer ($g_i^d > 0$) or farther ($g_i^d < 0$) growth stage) and how likely component $i$ occurs when the other factors are equal, respectively. To obtain p-values for each parameter, a permutation test [40] was used as 2,000 stimulations. All the components in $\sigma^{(k)}$ were converted to the 0-1 range as follows. We first interpret the relative abundance of each bacterial genus in the samples as the area close to the facility (No.31 in Fig. S6) (1) or 100 m away (No.31_100 m in Fig. S6) (0) states by setting a threshold value of 0.001. Then, we selected 51 families ( >1% as the maximum value of the bacterial population in all seasons). Consequently, we obtained the set of explanatory variables $\sigma^{(k)}$ with 51 components, which accompanies the environmental condition $\epsilon^{(k)}$ represents the status of the distance from the aquaculture facility and the seasonal stage.

### Structural equation modeling

Structural equation modeling (SEM) for confirmatory factor analysis (CFA) was performed using the R software package "lavaan" [41,42] as previously described [30]. In summary, the analysis code was based on the website (https://lavaan.ugent.be); since CFA requires a hypothesis, the bacteria selected in the association analysis, linear discriminant analysis, and energy landscape analysis were used as factors for the latent components of the data. The hypothesized model was statistically estimated by maximum likelihood (ML) parameter estimation with bootstrapping (n = 1000) using the functions 'lavaan' and 'sem'. The goodness of fit of the model was evaluated by chi-square p-value (p >0.05, not significant), comparative fit index (cfi/CFI) (>0.95), root mean square error of approximation (rsmea/RSMEA) (<0.05) and standardized mean residual (srmr/SRMR) (<0.08) [43]. When there were many candidate models, models with lower values of the Akaike Information Criterion (AIC) and smaller differences between the goodness-of-fit index (gfi/GFI) and the adjusted goodness-of-fit index (agfi/AGFI) (>0.95) were selected. Path diagrams for the superior models were visualized using the package "semPlot" in R software (Epskamp et al., 2019).

### Statistical causal inference

Three types of causal inference were performed with some modification, as previously reported[30]. Causal mediation analysis (CMA) was performed using the "mediation" [44] package in R software based on the tutorial website (https://rpubs.com/Momen/485122). First, each regression relationship with '~' in the selected model was assessed using the 'lm' function. Subsequently, the values of the causal relationships between bacterial candidates as mediators and outcomes were evaluated using the 'mediation' function. Finally, the estimated average causal mediation effect (ACME), average direct effect (ADE), and the proportion of the total effect through mediation were calculated by nonparametric bootstrapping ('boot=TRUE') with 'sims=1000' as the numbers of iterative calculations and quasi-Bayesian approximation ('boot=FALSE' as a command). Computable data are listed in Table S2. Direct LiNGAM, a linear non-Gaussian acyclic model [45], independent component analysis, and a non-Gaussian method for estimating causal structures were used to assess structural universality beyond the limited experimental data. DirectLiNGAM was established with Python code on the website (https://github.com/cdt15/lingam) (Python version 3.7.10). The p-values of the independence between the error variables were calculated simultaneously. Directed acyclic graphs (DAG) based on data calculated by Direct LiNGAM were visualized (Force Atlas with Noverlap) by Gephi. BayesLiNGAM[46], a Bayesian score-based approach that takes advantage of non-Gaussian distributions when evaluating linear acyclic causal models, was applied to the selected bacterial groups related to seagrass growth. BayesLiNGAM was established by the package 'fastICA' in R software based on the information from the specialized website prepared (https://www.cs.helsinki.fi/group/neuroinf/lingam/bayeslingam/). The package 'fastICA' in R was used according to other information on the website (https://cran.r-project.org/web/packages/fastICA). The seasonal DAGs calculated by BayesLiNGAM were visualized as networks in the package R 'igraph'.

### Other statistics

Regarding the statistical processing not mentioned above, the F test and Shapiro−Wilk normality test were used to determine whether a Gaussian or non-Gaussian distribution should be used. The Jarque-Bera test was also used in the 'tseries' library of R software. The unpaired $t$ test and the Wilcoxon test were performed as appropriate depending on the data. Significance was declared when $P < 0.05$, and a tendency was assumed at $0.05 \, P < 0.2$. These calculation data were prepared using Microsoft 365, R software, and Prism software (version 9.2.0)(GraphPad Software, LLC., USA). Data are presented as means ± SEs.

## Results

### Reinvestigation of public surveys of seagrass and seaweed distributions

Here, the distribution of seagrass along 49 coasts around Saiki city, Oita Prefecture (Fig. S2), was investigated by machine learning, as described in Fig. S1. As a result of the field survey, compared to information from 12 years ago, the exceptional appearance of seagrass was found around an onshore aquaculture facility was found (Fig. S3). This field survey revealed that the seagrass *Zostera* grew in the drainage channel and the coast surrounding the fishery farm (Fig. 1 and Movies S1). The relationship between seagrass overgrowth and regional characteristics was evaluated by association analysis (AA), an unsupervised machine learning method, with data from 49 coasts in the bay. The AA results (Figs. 2a-2c) estimated that seagrass (*Zostera*, Amamo as Japanese name) and seaweed (seaweed *Ecklonia*, Kajime as Japanese name; seaweed *Gelidiaceae*, Tengusa as Japanese name) grew near the coast without aquaculture facilities (AL in Fig. 2a), and seaweed *Sargassum* (Hondawara as Japanese name) was abundant there in 2004. It was estimated that seagrass (*Zostera*, Amamo as Japanese name) and seaweed (seaweed

*Monostroma*, Aosa as Japanese name; seaweed *Ecklonia*, Kajime as Japanese name; seaweed *Sargassum*, Hondawara as Japanese name; seaweed *Gelidiaceae*, Tengusa as Japanese name) did not grow near the coast with the facilities (AH in Fig. 2b), and two types of seaweed, *Ecklonia* and *Sargassum*, were also not abundant there in 2004. When the seagrass *Zostera* grew in 2016 (S in Fig. 2c), the AA results for the relationship among the growth, distance from the coast, and the presence of aquaculture facilities are shown. As a result, the flourishing of three types of seaweed (*Monostroma*, Aosa; *Ecklonia*, Kajime; *Sargassum*, Hondawara) was estimated at the same time as that of the seagrass *Zostera* grown in 2016. The three seaweeds did not flourish in 2004. In relation to the relationship with the distance from the coast, seagrass *Zostera* tended to grow around the area far from the coast (≧ at least 40 m as the distance from the shore). Two types of seaweed, *Ecklonia* and *Sargassum*, also tended to grow in the area far from the coast (≧ at least 20 m as the distance from the shore). The seaweed Monostroma did not have a substantial distance relationship (>1.2 as a lift value). These results suggested that aquaculture and the coast did not positively influence seagrass growth.

However, dense seagrass growth was observed around the coastal area in proximity to only one aquaculture facility (within 10 m from shore) after 2004 (No.31 in Figs. 1 and S3). The seagrass *Zostera* flourished in its drainage channel and the surrounding sea area around the aquaculture facility (Figs 1b, 1c, and S4). Therefore, we evaluated the characteristics of seagrass in this facility and the bottom sediment with the seagrass (Fig. 1d).

### *Biological and physicochemical conditions* of *the seagrass and its surroundings*

It was revealed that the seagrass *Zostera* was genetically *Z. marina,* as shown in the 'Method details'. The stable isotope carbon and nitrogen concentration in the seagrass itself was also confirmed (Table S1): the averages of $\delta^{13}C$, $\delta^{15}N$, total nitrogen, and total carbon, -11.9 ± 0.2, 7.1 ± 0.2, 1.9 ± 0.2, and 27.8 ± 1.2, respectively (the implications of these concentrations will be discussed later). A water quality analysis showed almost no difference in water quality between filtered groundwater and circulating seawater near the aquaculture facility (Table S2). The characteristics of the seagrass bottom sediment around this drainage channel (Movies S1, S2, and S4) and the bottom sediment in which no seagrass grew (approximately 100 m from the east coast of the drainage channel) (Movies S3 and S5) were investigated (Fig. S5a). In area No. 31 where seagrass grows (Fig. S5b), nine types of shellfish with different appearances were found in a radius of approximately 50 cm in a time of roughly 10 minutes (Fig. S5c). In area No.31_100 m, seaweed *Monostroma* were densely packed (Fig. S5d), and it was difficult to find shellfish. It took approximately 30 minutes to find three shellfish within a radius of roughly 2.5 m (Fig. S5e). The survey appeared to show the difference in abundance in the shellfish *Saxidomus purpurata* between the areas, although there are no strict statistical data. The physicochemical properties in both areas had different tendencies as follows: a significant increase in total carbon of the sediment and a decrease in the zinc content in the seagrass sediment (Figs. 2d-e). In addition, the levels of calcium, potassium, and magnesium tended to increase (Fig. S6). A decreasing trend of iron, manganese, phosphoric acid, and the humic index was confirmed (Figs. 2e and S6). Thus, these observations show differences in overall biological and physicochemical conditions, even though the two areas are very close and their distance from the shore is almost the same.

### *Seagrass sediment symbiotic bacterial candidates selected by machine learning*

Differences in bacterial diversity were irregularly observed between sediments without and with seagrass (Figs. S7-S9). These results appeared to indicate that the magnitude of variation depended on the timing of the sample collection. The abundance of minor bacterial genera (less than 1%) suggests that evaluation at the genus level is not always suitable (Fig. S9). Therefore, AA, as an unsupervised machine learning method, was used to select the bacteria related to the increase in seagrass: an increase in the order Clostridiales in the seagrass area and decreases in the order Myxococcales and Cytophagales were observed (Fig. S10a). In agreement, the abundances of the families Lachnospiraceae, Ruminococcaceae, and Veillonellaceae, belonging to the order Clostridiales, are shown there (Figs. 3a and 3b). In addition, the families Bacteroidaceae, Desulfobulbaceae, Rhodobacteraceae, and Rhizobiaceae also had an increasing tendency. However, the families of Cryomorphaceae belonging to the order Flavobacteriales, Saprospiraceae belonging to the order Saprospirales, and Flammeovirgaceae belonging to the order Cytophagales had a decreasing tendency (Fig. 3c). Next, LDA analysis (Figs. S10b and 10c), a supervised machine learning method, succeeded in extracting bacteria with significant differences among the results calculated by the AA. A considerable increase in the family Desulfobulbaceae was observed in seagrass sediment (Fig. S10b). A decrease in the order Flavobacteriales was observed (Fig. S10b). These cladograms were shown in Fig.S10c. When the seasonal variation of detected bacteria (Figs. S10a and S10b) was evaluated, all families belonging to the order Cytophagales, Flavobacteriales, and Myxococcales characteristic in the area without seagrass (Fig. S10d). The relative abundance was reduced in the area with seagrass. In addition, the upward and downward trend of the AA-detected bacterial families (Fig. 3a) was confirmed beyond seasons despite random sampling (Fig. S11).

The relationship between the distance from the aquaculture facilities and the seasons was assessed by ELA (Fig. S12). The AA- and LDA- detected bacterial families were selected by spread around the 0 value of the Y-axis (Effects of the season). Further, the analysis of the energy landscape succeeded in detecting the effects of season and distance of the compositions of the bacterial community related to the growth of seagrass (Fig. 4). Two types of stable states (Cs0 and Cs1) were identified by a calculation based on the seasons in the landscape were identified (Fig. 4a). Five stable states (C0_1, C0_2, C0_3, C1_2, and C1_3) were identified by calculation based on the distance in the landscape (Fig. 4b). These results suggest that the effect of distance is stronger than the seasonal effect. Therefore, the proximity to the aquaculture facility is essential. In other words, it is contrary to the conclusion of the AA analysis that the seagrass grows farther from shore (Fig.

2c). These ELA results indicate that the characteristics of Area 31, where seagrass growth was observed near the aquaculture facility, deserve attention.

Next, the clusters of these stable states were statistically compared, and the statistical importance of p1, n2, and n3 was emphasized (Figs. S13). These three clusters were composed of the following groups of bacteria: p1, Desulfarculaceae, Lachnospiraceae, Campylobacteraceae, Enterobacteriaceae, Bacteroidaceae, S24-7, Bifidobacteriaceae, Streptoococcaceae, Veillonellaceae, Rhizobiaceae, Helicobacteraceae, [Acidaminobacteraceae], and Coxiellaceae (Fig. 4c); n2, Nitriliruptoraceae, Euzebyaceae, Nocardioidaceae, Hyphomonadaceae, Rhodothermaceae, Rhodospirillaceae, and Ellin6075 (Fig. 4d); and n3, Rhodobacteraceae, Thiotrichaceae, Desulfobulbaceae, Desulfobacteraceae, and Flavobacteriaceae (Fig. 4e). Of these three clusters, two clusters (Figs. 4c and 4e) contained AA- and LDA- selected bacteria.

Furthermore, seasonal changes in these bacteria selected by ELA are shown in Fig. 5. However, it has already been shown that the proximity to this aquaculture facility has a greater impact (Fig. 4a and 4b). Consequently, ELA may estimate that alteration of the balance of these bacterial groups is involved in a geographical and seasonal relationship for the growth of seagrass around the aquaculture facility. Therefore, flexible changes in the composition of these bacterial groups may control seagrass overgrowth in a balanced manner depending on geographical conditions and seasonal fluctuations. Among these bacterial groups selected by machine learning, the bacteria that had increasing trends in the seagrass overgrowth area beyond seasonal conditions and the ones that showed significant differences in common in the three analyses should be notable.

### *Candidates for the optimal symbiotic bacterial group of seagrass sediments estimated by SEM and causal inference*

Structural equation modeling and statistical causal inference for seagrass growth were performed using bacteria selected by machine learning and ones with high abundance ratios (Figs. 3-5). For the latter, the family Rodobacteraceae (Figs. 4e and 5) had involved in the ELA interaction network n3 with Desulfobulbaceae, which was selected by three machine learning methods. The family Rodobacteraceae is relatively abundant in the environment as a photosynthetic bacterium. It was present in more than 5% of seagrass flourishing areas (area No.31), although there is no statistically significant difference between the two areas (p=0.2215)(Fig. S14a).

Structural equation modeling (SEM) linked to seagrass presence was performed (Table S3). Five types of relatively ideal structural model candidates with orders, families, and/or genera were formulated. These goodness-of-fit indices were appropriate. Since the success rate of the bootstrapping calculation was also low (39 draws per 1000 times), model No.4 was taken as a slightly inferior model. Since the underlined AGFI value was low (preferable > AGFI 0.95), model No.5 was considered somewhat flawed. It is difficult to select the optimal model from No_1 to No.3. The AIC value was the lowest for the model No. 3, while the slightest difference between the GFI and AGFI values occurred in model No. 1. The lowest SRMR value was the model No. 2, but the success rate of the bootstrapping calculation was the lowest (478 draws per 1000 times) among the three models. The No.1 model, which had the highest chisq p value of 0.928 without the underlined values, was composed of the family Desulfobulbaceae selected with high reproducibility by machine learning. The genera belonging to the families comprising the No.1 model behaved similarly to the families, although there were some exceptions (Fig. S14b). In addition, the model (No. 6) for the genus was inferior in value to model No.1. These calculation results do not deny the importance of the bacteria selected by machine learning but highlight the candidates of the bacterial group that are intensely estimated by covariance structure analysis. Thus, the No.1 model with the families Desulfobulbaceae, Lachnospiraceae, Rhodobacteraceae, and Ruminococcaceae was visualized as an optimal structural model (Fig. 6a). Furthermore, the causal mediation analysis (CMA) in Fig. 6a confirmed the importance of the optimal model as a group rather than a single prominent bacterium (Table S4).

The experimental data were sampled randomly across seasons. Therefore, universality is assumed high, but SEM uses the maximum likelihood method based on a Gaussian distribution. To further generalize the phenomenon, it is necessary to analyze the assumption of a non-normal distribution in nature. The distribution test with the Shapiro−Wilk test and the Jarque Bera test partly showed a non-Gaussian distribution (Fig. S15). Therefore, assuming a non-Gaussian distribution, the values of the symbiotic causal structure of these bacteria were estimated by DirectLiNGAM, a linear non-Gaussian acyclic model as a statistical causal inference analysis (Fig. 6b). Assuming an independent component, DirectLiNGAM estimated a strong relationship among the families Lachnospiraceae and Ruminococcaceae belonging to the order Clostridiales, the family Desulfobulbaceae, and family Rhodobacteraceae, in the sediment with seagrass.

Thus, the model with the families Desulfobulbaceae, Lachnospiraceae, Rhodobacteraceae, and Ruminococcaceae can be manifested under both conditions with Gaussian distributions and non-Gaussian distributions, depending on the environmental conditions (Figs. 5c). At least, they should be regarded as influential candidates of symbiotic bacterial groups with a crucial relationship that helps seagrass growth more efficiently near the onshore aquaculture facility.

### Discussion

The exceptional growth of seagrass around the onshore aquaculture facility in this study is an essential finding for sustainable fishery industry. Therefore, the candidates of the symbiotic bacterial group involved in these seagrass overgrowths could be estimated by machine learning, statistical structure equation modeling, and causal inference. In addition, the concentrations of stable isotope carbon and nitrogen in the seagrass itself tended to be slightly higher (Table S1) than the available concentrations in the bottom sediments of the sea, the areas of the seagrass and phytoplankton[47]. In particular, *Saxidomus purpurata* has increased here (Fig. S6), although there are no strict statistical data. In general, shellfish contain many zinc, iron, etc.[48,49], so it is interesting that these chemical components were reduced in the bottom sediment where seagrass is located. Therefore, these phenomena were not inconsistent with the concept of past fundamental findings that the symbiosis of seagrass and shellfish is essential in the seagrass ecosystem in terms of environmental purification. [50].

The characteristics of AA- and/or LDA-detected bacteria can be understood as follows: symbiotic bacteria in seagrass bottom sediment have been investigated under various environmental conditions[51,52]. As one of the trends, common terms with the intestinal flora of animals and the symbiotic bacteria of terrestrial plants on land have also been confirmed. The families Bacteroidaceae, Ruminococcaceae, and Lachnospiraceae are producers of short-chain fatty acids, which play a role in biological defenses [53-55], and the family Rhizobiaceae contains nitrogen-fixing bacteria [56]. The order Flavobacteriales is known to decrease around seagrass[14,57]. It should be noted that the Flavobacteriales order contains bacteria pathogenic to fish [14,57] and that the Fusobacteriaceae family has pathogens strongly associated with human cancer [58]. The family Saprospiraceae is known to be abundant in sewage [59,60]. Therefore, the poor abundance in seagrass sediment may be caused by water quality purification by seagrass with *Saxidomus purpurata (Fig. S6) as a shellfish, as previously described* [50]. In addition, purification may be involved in the decrease in the family Saprospiraceae.

In particular, bacterial candidates in the optimal structural equation model in seagrass overgrowth are expected to play an important role in seagrass overgrowth. The family Desulfobulbaceae is a cable bacteria in sea sediments [22,61-63]. It has been pointed out that this family may also be involved in the nitrogen cycle [64,65], and such cable bacteria have been reported worldwide [23]. The abundance of the family Desulfobulbaceae, as symbiotic seagrass bacteria, has been also confirmed in the sediment of several seagrasses, *Thalassia hemprichii* [66] and *Halophila ovalis* [52], in different regions. Furthermore, the family Rhodobacteraceae, which has been detected in ocean sediments worldwide [24], although not always along the coast, is an essential bacterial family for environmental protection in the sea. Rhodobacteraceae is a symbiont of seagrass and $CO_2$ fixation [67]. Furthermore, genera belonging to Rhodobacteraceae play crucial roles: a *Paracoccus* strain produces astaxanthin with high antioxidant activity[68]; a *Phaeobacter* strain enhances the ability of seabass larvae to survive infections caused by a mild pathogen [69]. Therefore, the calculation of the structural equations with the genera belonging to the family Rhodobacteraceae formulated model No. 5 (Table S3). The underlined value (AGFI 0.908) was less than 0.95 (preferable > AGFI 0.95). The model may have these essential meanings as a functional bacterial group, even if the value was not optimally fitted

Our computational models will help guide the developmental study of blue carbon and aquatic ecosystem services, although this result may be confined to the local area. It is necessary to devise ways to increase seagrass density further, as seagrass density is extremely important for fish survival [70]. This study shows the relationship between seagrass overgrowth in the vicinity of this aquaculture facility and symbiotic bacterial groups on both models of a Gaussian and non-Gaussian distribution. The goodness-of-fit by calculation hypothesized that these distributions and the results of the referred searches might be assumed to indicate that the estimated seagrass symbiotic bacterial group candidates have a relatively high universality. In general, microbial communities in the environment have a complex influence on each other. Therefore, the causal relationship in this study still needs to be carefully discussed since other bacteria linked with these bacteria have also been confirmed by machine learning results. However, they are certainly a group with a strong relationship in terms of calculation. In addition, even considering the functional background of the estimated bacterial group described above, it is recalled that it may be involved in changes in metabolites near aquaculture facilities. ELA analysis showed that seagrass growth near the aquaculture facility was more affected by distance than by the seasons (Fig. 4). However, seagrass overgrowth fluctuates seasonally. Therefore, the seasonal behavior of these candidates of the optimal SEM bacterial group (Fig.6) was verified. The potential seasonal causalities within the optimal SEM bacterial group were statistically inferred by BayesLiNGAM, which can assume a causal relationship among the selected components (Fig. S16). The values of each spatial-temporal relationship between the bacteria were interrelated with fluctuations, as well as with the importance of the existence of the bacterial group and the seagrass as a whole. In ecosystems, the concept of fluctuations in the interactions between complex factors has been suggested [30,71-74]. Previous reports analyzing the behavior of chemicals in seaweed have also suggested a similar concept of fluctuations [75,76]. Here spatial and temporal (seasonal) fluctuations in seagrass sediment symbiotic bacteria could be statistically clarified.

Crucially, this aquaculture facility relied on exposure to fermented feed produced by recycling technology. Generally, recycled fermented feed is unstable due to various factors, such as fermentation bacteria, moisture levels, and other conditions of the fermentation process. However, nitrogen and phosphorus are retained and effectively used if the fermentation product is circulated. It is known that the agricultural environment amended by controlled compost may be improved [77]. A typical bacterium commonly found in compost is *Bacillus* [78,79]. *Bacillus* is one of the bacterial genera found in the soil of high-temperature environments. *Bacillus*, one of the Bacillaceae families, was reported to be a seagrass symbiont [80] Symbiotic microorganisms facilitate seagrass growth, and in particular, plant growth-promoting bacteria (PGPB) for terrestrial crops were implied to play a potential role in seagrass growth.

Notably, in the facility with seagrass growth, a recycled compost consisting of fermented marine animal resources at above 70 ° C with a thermophilic *Bacillaceae*[81] was continuously used as a feed additive after 2004 (Fig S1). The effects of exposure to the compost used as a fertilizer substitute and a feed additive on plants and animals have been evaluated. This compost comprises PGPB belonging to the thermophilic *Bacillaceae* family with antifungal activity [81]. The compost suppresses nitrate accumulation in plants[82] and promotes nitrogen fixation ($N_2$) based on alteration of soil bacterial flora, as a recent report [83]. In addition, the effect on animals has also been confirmed. The oral administration of this compost improved fertility and quality and reduced disease in rodent models, fish (flatfish, carp, and Red seabream), and livestock animals (chicken, pig, cattle)[54,84-89]. Compost-derived *Bacillaceae* also altered the animal bacterial flora, such as the decrease in the population of fecal methanogens that produce the greenhouse gas $CH_4$ in calves[54]. Furthermore, the compost itself appears to modulate the function of intestinal metabolism of animals [54,84-89], showing the potential to reduce the use of antibiotics [89]. A better outcome to reducing antibiotic use is expected when combined with the effect of lowering pathogens caused by seagrass [14]. In these respects, *Bacillaceae* fermented feed used in the seagrass-flourishing aquaculture facility may be effective in developing seagrass growing technology and environmental conservation. However, the detailed mechanism of action is unknown. It is well known that the bacterial flora of sea sediments is affected by the gut flora of symbionts [90]. The results of this study may also be influenced by compost itself, which modifies the gut bacteria of cultured fish in onshore aquaculture facilities and/or the modification of intestinal bacteria of organisms that live in symbiosis with seagrass. Furthermore, plankton is also crucial in marine ecosystems [91], although we have not examined it in relation to these conditions in this study. Further research on these in its entire ecosystem [92] will be necessary for the future.

Generally, the fishery industry needs to grow to overcome future food shortages [4], and it is necessary to consider the balance between fish quality and environmental load. The exceptional growth of seagrass around the farm in this study will impact global ecological conservation and efforts to overcome food shortages. Thus, these results are computationally and initially estimate a potential ideal "regime shift" of the seagrass ecosystem in the fishery farm. Our observations can provide an essential hint for the future food industry and environmental protection, as research on the ideal symbiotic microbial community of seagrass and/or seaweed, the coexistence of aquaculture and recycling technologies, and their quality.

**Acknowledgments** The authors especially thank Mr. Yasuyuki Miyoshi (Oita Prefecture), Fisheries Promotion Division (Oita Prefecture), Oita Institute of Marine and Fisheries Science (Oita Prefecture), and Oita Prefecture for viewing and citing public survey reports (2004 and 2016), Ms. Junko Ichikawa (Shoko Science Co., Ltd.), Ms. Risa Monna (Reso Co., Ltd.), Ms.Akiyo Muto, and Mr. Takashi Kobayashi (Agrokanesho Co., Ltd.) for special technical support and advice, and Hiroshi Hirakawa for filming the videos and taking the photographs.

**Author contributions** Conceptualization: H.M., N.K., H.M.; Methodology: H.M., N.K., A.K., S.M.,K.S.,H.M.,H.M., J.K.; Materials: U.K. and H.M. Investigation: H.M., A.K., W.S., S.M., K.S., M.M., N.T., T.N., C.I., A.T., C.S., T.K., T.S. H.O., J.K.; Visualization: H.M., N.K., K.S., N.T., J.K.; Project administration: H.M., N.K., Supervision: H.M., H.O., J.K.; Writing – original draft: H.M., J.K.; Writing – review & editing: H.M., A.K., W.S., S.M., K.S., T.K., S.W., T.S. H.M., H.O., J.K.

**Data availability** All data are available in the main text, figures, supplementary materials, movies, and data set files. In addition, the data sets presented in this study can be found as Excel files (named Data S1 and S2) in the supplementary information. Furthermore, all 16S rRNA gene datasets are deposited in the DDBJ Sequence Read Archive and can be found under accession numbers: LC636204 and PRJDB12183. Moreover, the R protocols for association analysis used in this study were described on the following websites (named Market Basket Analysis): http://dmar.riken.jp/Rscripts/ and http://dmar.riken.jp/NMRinformatics/.

**Competing interests** The authors declare no competing interests.

**Figure legends**

**Fig. 1**
Target regions of Japan and sampling sites in this study. (a) The study area was one of the inlets into which the Kuroshio current enters, and there is a high concentration of onshore aquaculture facilities for flounder in the sea. Area No. 31, circled in red, indicates an onshore aquaculture facility where the seagrass *Zostera marina*, grows exceptionally well. Photographs of (b) the seagrass in the drainage channel there, (c) the seagrass on the beach near the aquaculture facility, and (d) the white circle site where the bottom sediment sample was collected. These photographs were taken during low tide in June 2017.

**Fig. 2**
Verification of the geographical and physical environmental conditions in which seagrass and seaweed flourish. Vegetation association networks in areas (a) without and (b) with the presence of onshore aquaculture facilities, (c) factors related to the emergence of the seagrass *Zostera marina* in 2016, and (d) total carbon, zinc, and seagrass *Zostera* in the sediment. The abbreviations indicate as follows: AL, absence of aquaculture facilities; AH, presence of aquaculture facilities; E, seaweed *Ecklonia* (Kajime); G, seaweed *Gelidiaceae* (Tengusa); M, seaweed *Monostroma* (Aosa); Sar, seaweed *Sargassum* (Hondawara); Z, seagrass *Zostera* (Amamo)(terms in above parentheses indicate Japanese names); S, the presence of seagrass *Zostera* around aquaculture facilities in 2016; D, distance from the shore (followed by a number that indicates the distance from the coast); Be, data from 2004 (means Before); and Af, data from 2016 (means After). Furthermore, the trends of the high (_H) (red) or low (_L) (blue) levels based on the mediation values of all the data from the targeted components are shown (> 1.2 as lift values, n=49) (d) A selected association network of physicochemical factors of the sediments in areas with and without seagrass (around the area No.31 and the area No.31_100m in Fig. S6). The trends of high (_H) or low (_L) levels based on the mediation values of the whole data of targeted components are shown (> 1.5 as lift values; > 0.8 as confidence values, n=4) . (e) Estimation plots of total carbon, nitrogen, iron, and zinc in the sediment (n=4).

**Fig. 3**
(a) Association networks of bacterial families related to the presence of the seagrass *Zostera marina* (S in the center circle). The trends of high (_H) or low (_L) levels based on the mediation values of all data of the targeted components are shown (> 1.5 as lift values; > 0.8 as confidence values). (b)(c) shows the population of bacterial families selected by association analysis (>1.5 as lift values) in Fig. 3a (n=9). The p values less than 0.1 are shown on the graph.

**Fig. 4**
The entire energy landscape linked with the (a) season detected in this study and (b) distance away from the targeted onshore aquaculture facility (near area No.31 in Fig. S5). (a) The axis formed the energy landscape with compositional energy, community state, and season. Cs0 and Cs1 show the labels of seasonal stable states. Labels with seasonal values were established as follows: $g^s=0$, February; $g^s=0.325$, November; and $g^s=1$, June. (b) The axis formed the energy landscape with compositional energy, community state, and distance. Distance values were shown as $g^d=0$, area No.31_100m; and $g^d=1$, area No.31 in Fig. S5. The significant interaction network of (c), (d), and (e) shows p1, n2, and n3 in Fig. S13, respectively. The families with green letters offer the components selected by the AA. The underlined families show the component selected by LDA.

**Fig. 5**
Seasonal comparison heatmap with bacterial families in clusters selected by ELA (Fig. 4). The clusters in p1 (Fig. 4c), n2 (Fig. 4d), and n3 (Fig. 4e) show each group with a significant value (p<0.05) in a stable state of Figs. 4 and S13. The families surrounded by the green and red squares show the components selected by AA and LDA, respectively. The asterisks indicate the families with the significant values as each of the whole data (p<0.05).

**Fig. 6**
Visualization of SEM and causal relationship in the symbiotic bacterial group of seagrass sediment. (a) Optimal bacteria SEM is reported in sediment with seagrass standardized β coefficients. The abbreviations are as follows: Dsulf, family Desulfobulbaceae; Lachn, family Lachnospiraceae; Rhodo, family Rhodobacteraceae; Rumin, family Ruminococcaceae; Chisq, chi-square $\chi^2$; p value, p values (Chi-square); CFI, comparative fit index; TLI, Tucker–Lewis index; RMSEA, root mean square error of approximation; SRMR, standardized root mean residuals; GFI, goodness-of-fit index; and AGFI, adjusted goodness-of-fit index. Green and red indicate positive and negative effects, respectively. The selected fit indices are shown within the path and in Table S2. (b) The directed acyclic graph (DAG) was calculated by DirectLiNGAM (see Fig. S14) based on the optimal SEM shown in Fig. 6a. Each number in the DAG shows the calculated causal value.

**Fig. 7**
The putative role of the optimal symbiotic bacterial group of seagrass predicted in this study.

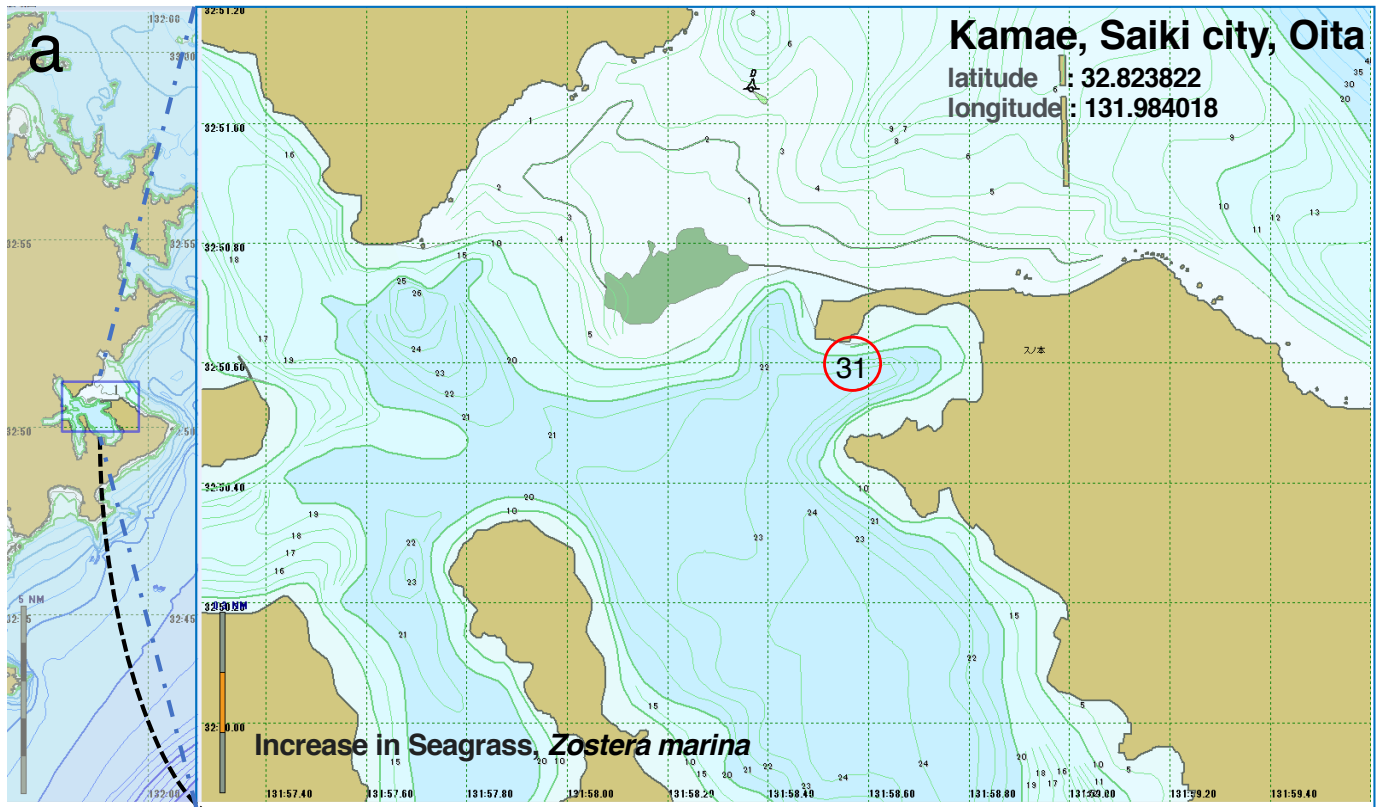
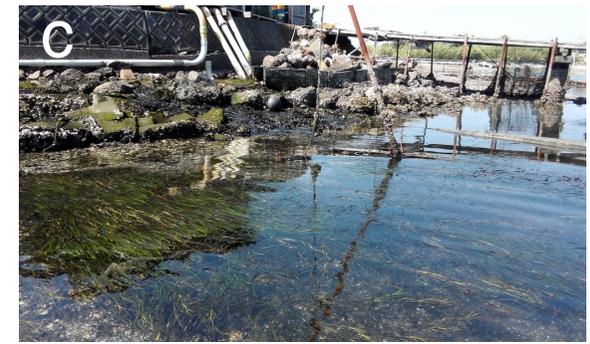
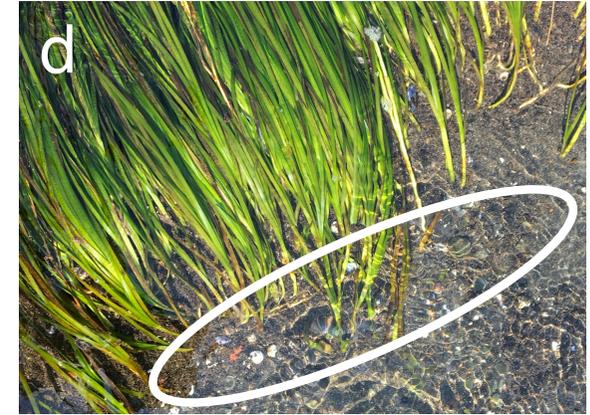
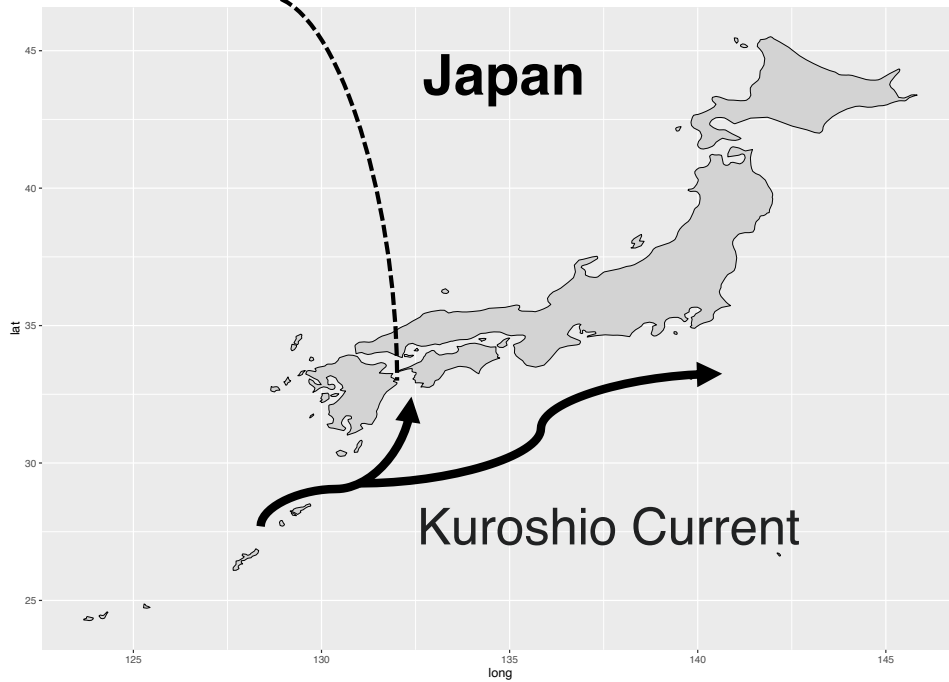
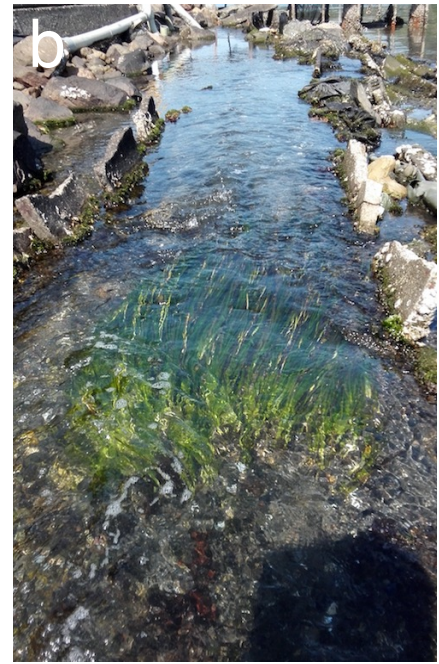

**Fig. 1**

Target regions of Japan and sampling sites in this study. (a) The study area was one of the inlets into which the Kuroshio current enters, and there is a high concentration of onshore aquaculture facilities for flounder in the sea. Area No. 31, circled in red, indicates an onshore aquaculture facility where the seagrass *Zostera marina*, grows exceptionally well. Photographs of (b) the seagrass in the drainage channel there, (c) the seagrass on the beach near the aquaculture facility, and (d) the white circle site where the bottom sediment sample was collected. These photographs were taken during low tide in June 2017.

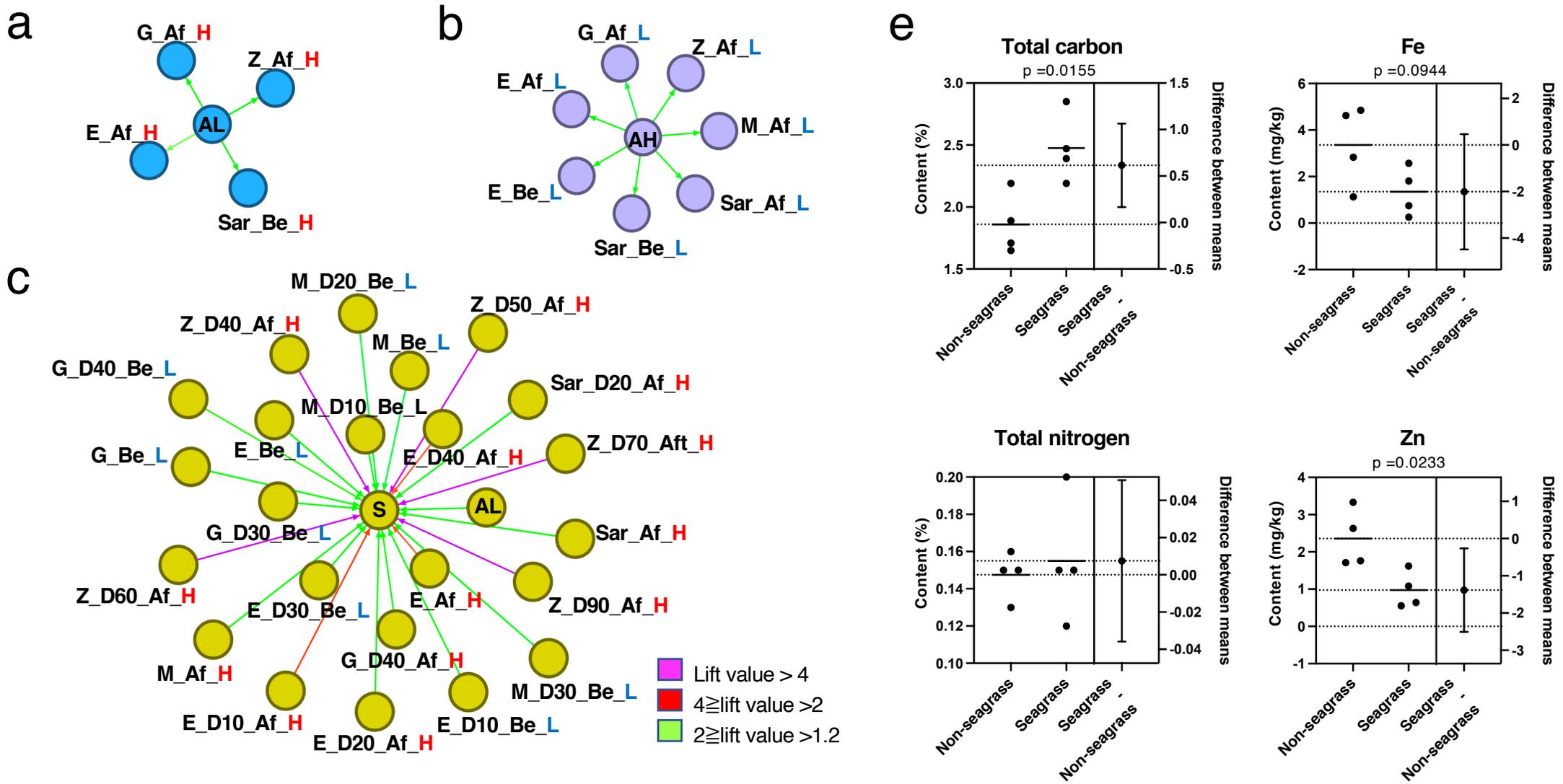

**Fig. 2**
Verification of the geographical and physical environmental conditions in which seagrass and seaweed flourish. Vegetation association networks in areas (a) without and (b) with the presence of onshore aquaculture facilities, (c) factors related to the emergence of the seagrass *Zostera marina* in 2016, and (d) total carbon, zinc, and seagrass *Zostera* in the sediment. The abbreviations indicate as follows: AL, absence of aquaculture facilities; AH, presence of aquaculture facilities; E, seaweed *Ecklonia* (Kajime); G, seaweed *Gelidiaceae* (Tengusa); M, seaweed *Monostroma* (Aosa); Sar, seaweed *Sargassum* (Hondawara); Z, seagrass *Zostera* (Amamo)(terms in above parentheses indicate Japanese names); S, the presence of seagrass *Zostera* around aquaculture facilities in 2016; D, distance from the shore (followed by a number that indicates the distance from the coast); Be, data from 2004 (means Before); and Af, data from 2016 (means After). Furthermore, the trends of the high (_H) (red) or low (_L) (blue) levels based on the mediation values of all the data from the targeted components are shown (> 1.2 as lift values, n=49) (d) A selected association network of physicochemical factors of the sediments in areas with and without seagrass (around the area No.31 and the area No.31_100m in Fig. S6 (> 1.5 as lift values; > 0.8 as confidence values, n=4). (e) Estimation plots of total carbon, nitrogen, iron, and zinc in the sediment (n=4).

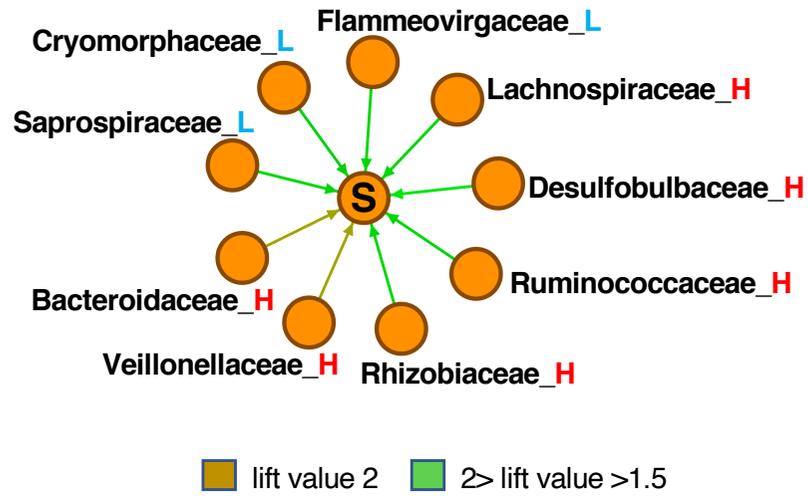
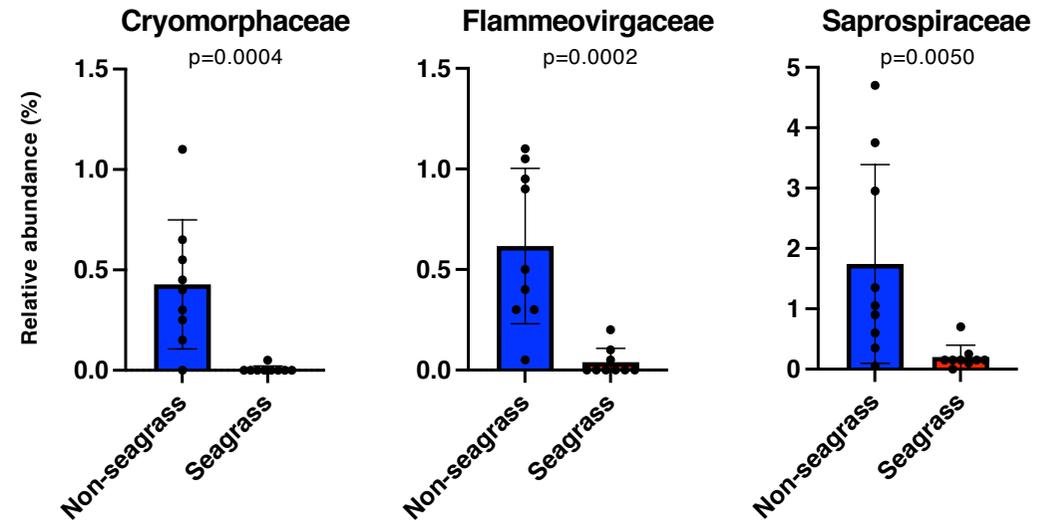
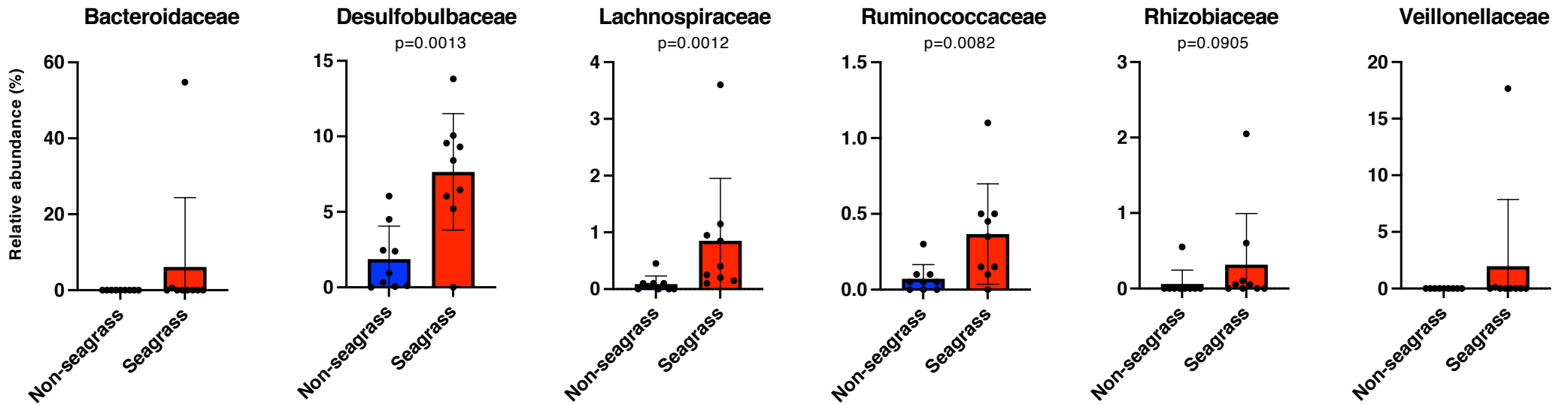

**Fig. 3**
(a) Association networks of bacterial families related to the presence of the seagrass *Zostera marina* (S in the center circle). The trends of high (_H) or low (_L) levels based on the mediation values of all data of the targeted components are shown (> 1.5 as lift values; > 0.8 as confidence values). (b)(c) shows the population of bacterial families selected by association analysis (>1.5 as lift values) in Fig. 3a. The p values less than 0.1 are shown on the graph.

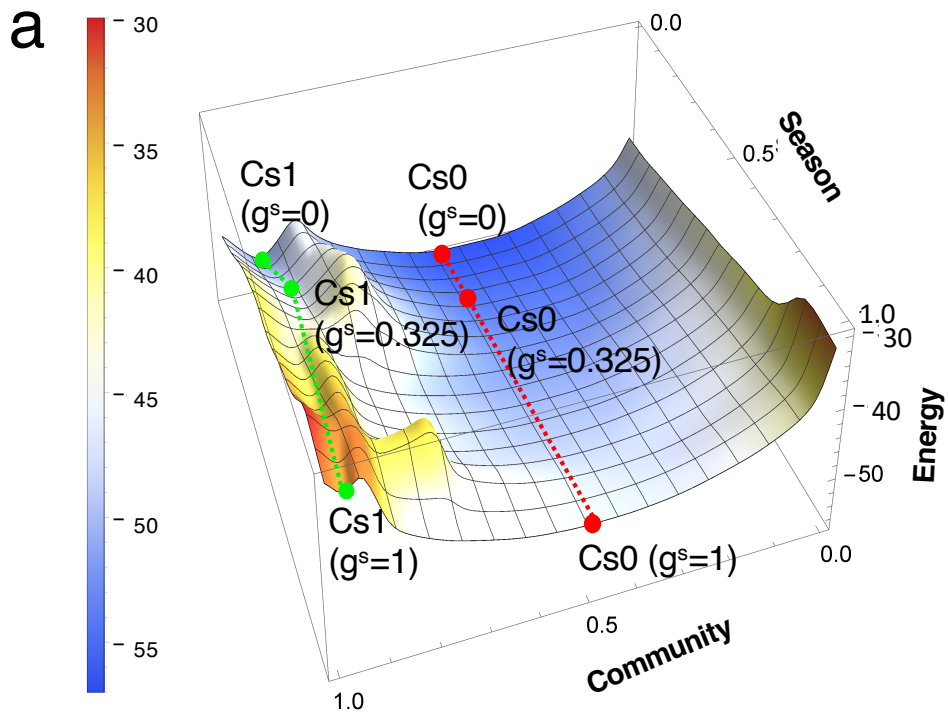
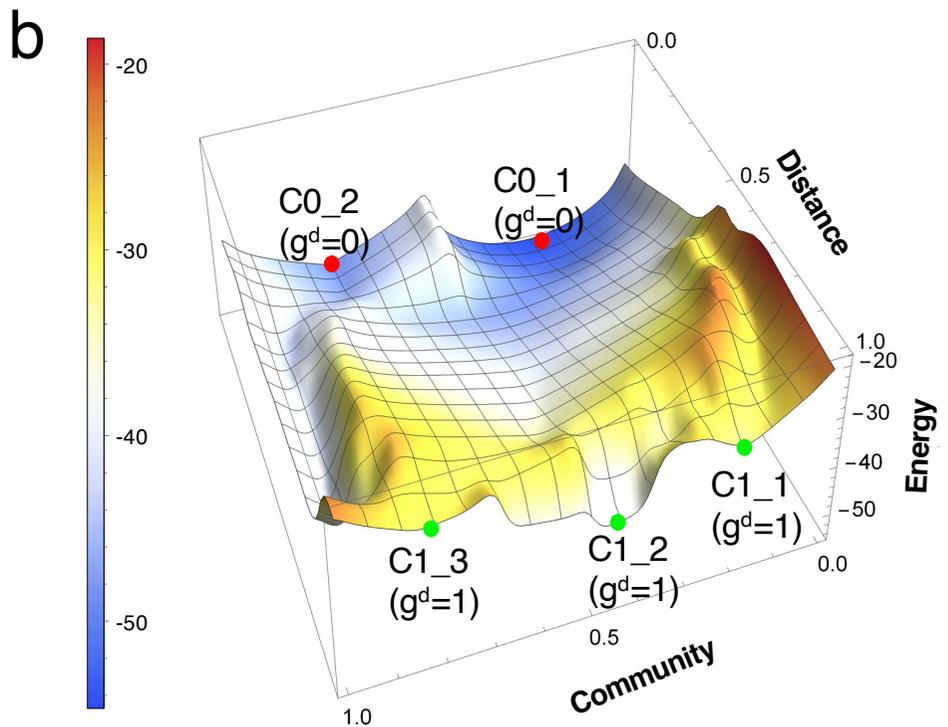
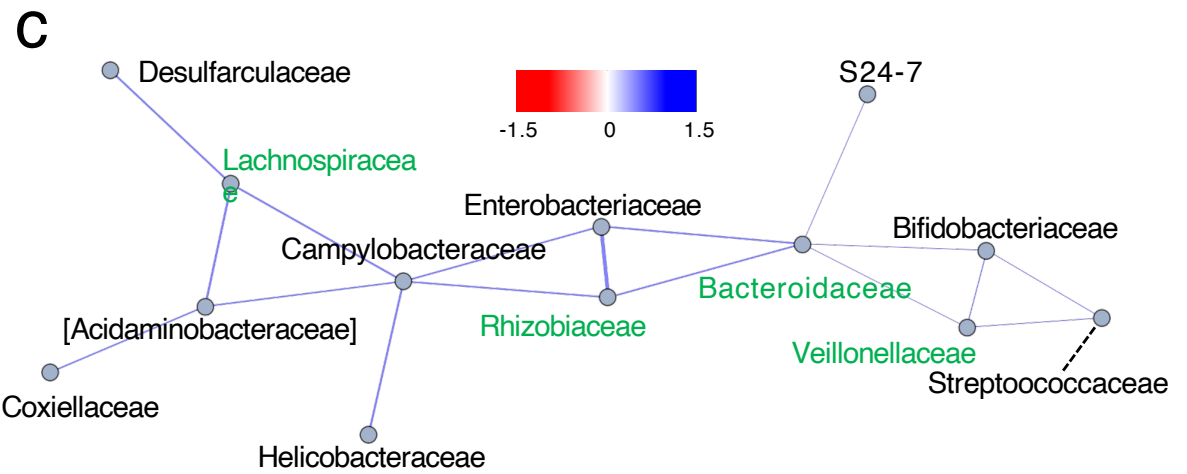
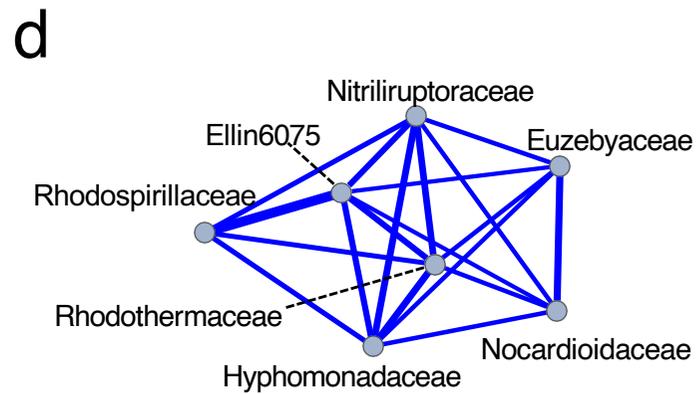
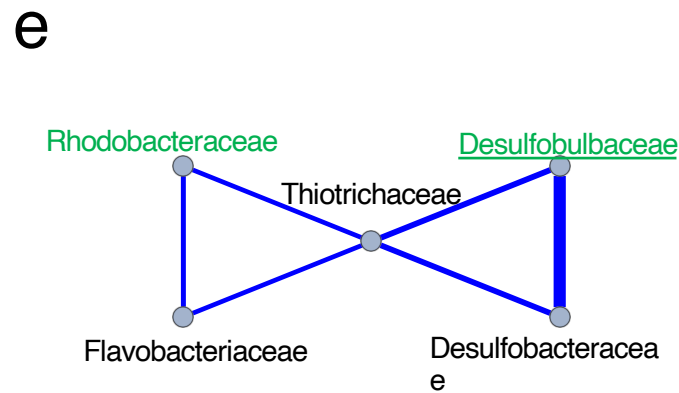

**Fig. 4**
The entire energy landscape linked with the (a) season detected in this study and (b) distance away from the targeted onshore aquaculture facility (near area No.31 in Fig. S5). (a) The axis formed the energy landscape with compositional energy, community state, and season. Cs0 and Cs1 show the labels of seasonal stable states. Labels with seasonal values were established as follows: $g^s=0$, February; $g^s=0.325$, November; and $g^s=1$, June. (b) The axis formed the energy landscape with compositional energy, community state, and distance. Distance values were shown as $g^d=0$, area No.31_100m; and $g^d=1$, area No.31 in Fig. S5. The significant interaction network of (c), (d), and (e) shows p1, n2, and n3 in Fig. S13, respectively. The families with green letters offer the components selected by the AA. The underlined families show the component selected by LDA.

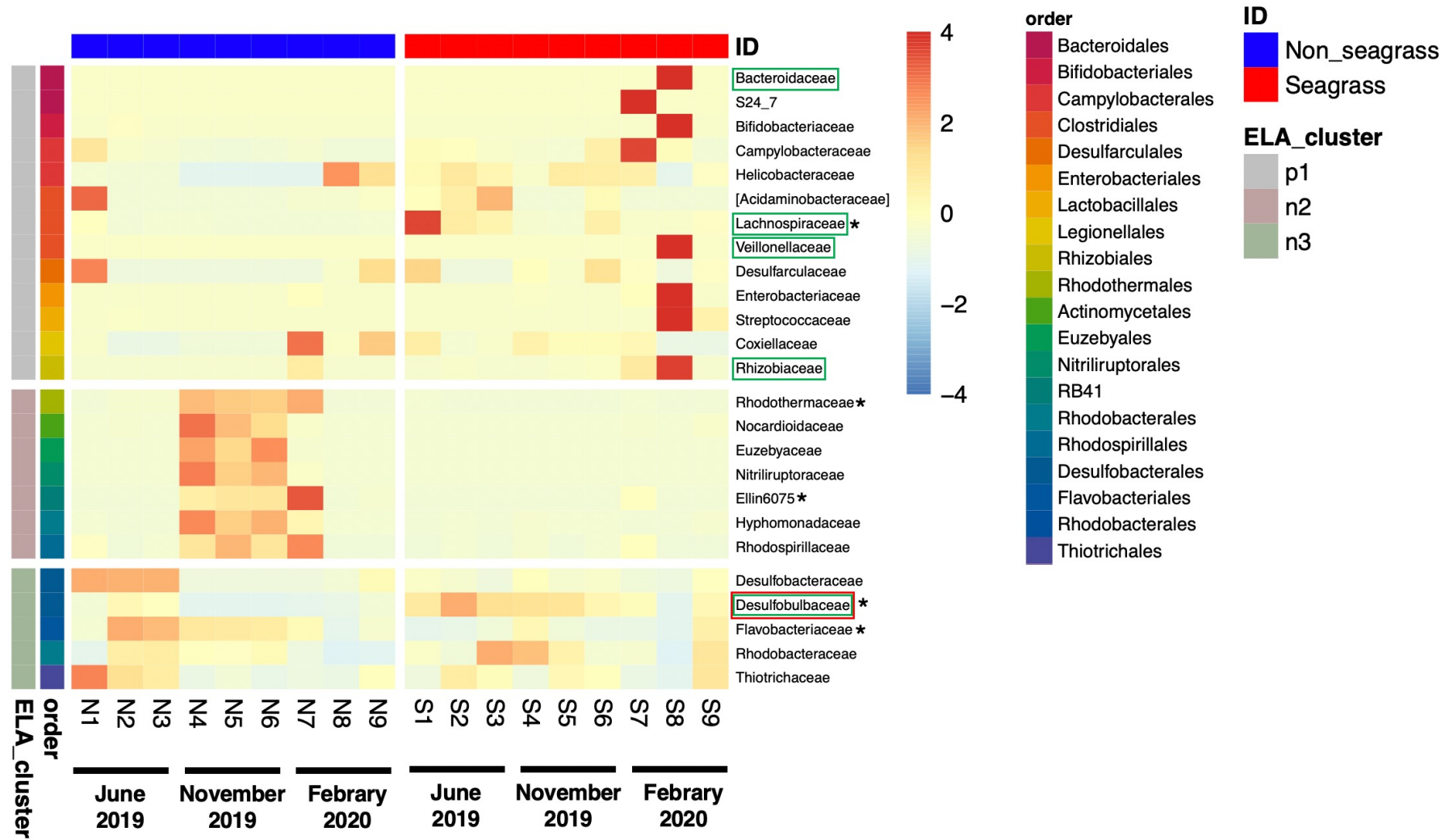

**Fig. 5**
Seasonal comparison heatmap with bacterial families in clusters selected by ELA (Fig. 4). The p1 (Fig. 4c), n2 (Fig. 4d), and n3 (Fig. 4e) show each Seasonal comparison heatmap with bacterial families in clusters selected by ELA (Fig. 4). The clusters in p1 (Fig. 4c), n2 (Fig. 4d), and n3 (Fig. 4e) show each group with a significant value (p<0.05) in a stable state of Figs. 4 and S13. The families surrounded by the green and red squares show the components selected by AA and LDA, respectively. The asterisks indicate the families with the significant values as each of the whole data (p<0.05).

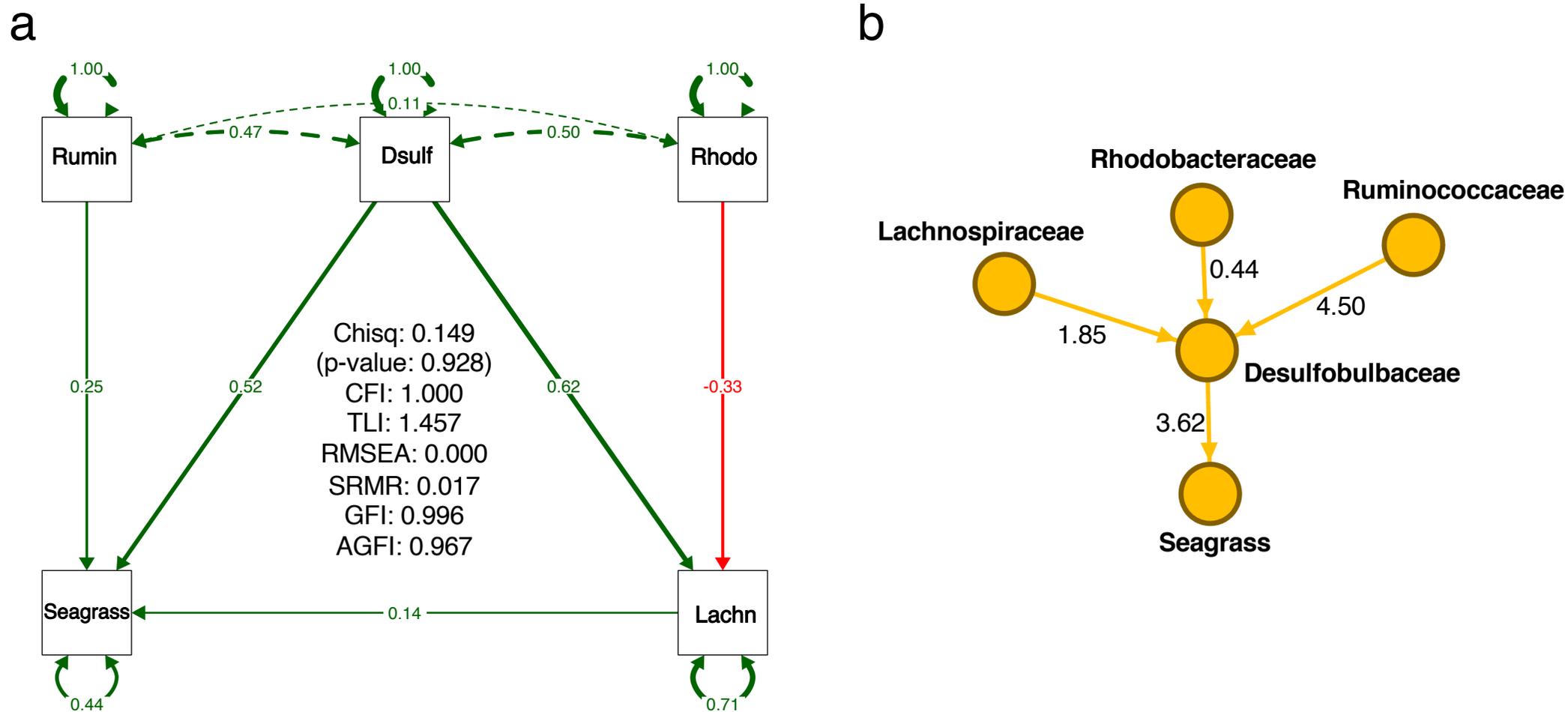

**Fig. 6**

Visualization of SEM and causal relationship in the symbiotic bacterial group of seagrass sediment. (a) Optimal bacteria SEM is reported in sediment with seagrass standardized β coefficients. The abbreviations are as follows: Dsulf, family Desulfobulbaceae; Lachn, family Lachnospiraceae; Rhodo, family Rhodobacteraceae; Rumin, family Ruminococcaceae; Chisq, chi-square $\chi^2$; p value, p values (Chi-square); CFI, comparative fit index; TLI, Tucker–Lewis index; RMSEA, root mean square error of approximation; SRMR, standardized root mean residuals; GFI, goodness-of-fit index; and AGFI, adjusted goodness-of-fit index. Green and red indicate positive and negative effects, respectively. The selected fit indices are shown within the path and in Table S2. (b) The directed acyclic graph (DAG) was calculated by DirectLiNGAM (see Fig. S14) based on the optimal SEM shown in Fig. 6a. Each number in the DAG shows the calculated causal value.

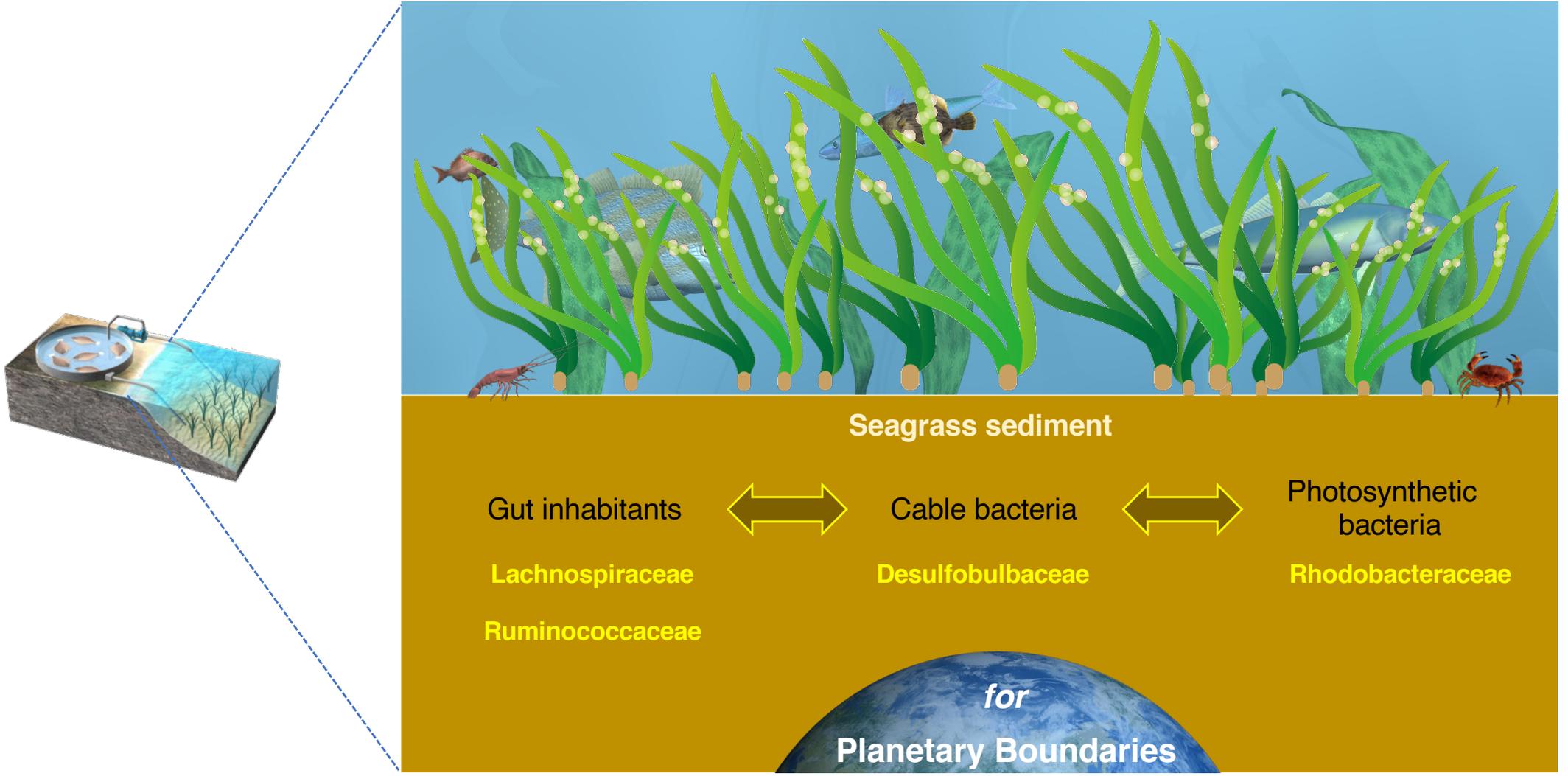

**Fig. 7**
The putative role of the optimal symbiotic bacterial group of seagrass predicted in this study.

# Supplementary Information

# A potential basic symbiotic bacterial structure that contributes to seagrass growth downstream of onshore aquaculture

*Running head*:     Symbiotic Bacteria of Seagrass and Aquaculture

Hirokuni Miyamoto*[1,2,3,4], Nobuhiro Kawachi[5], Atsushi Kurotani[6], Shigeharu Moriya[7], Wataru Suda[2], Kenta Suzuki[8], Makiko Matsuura[1,4], Naoko Tsuji[4], Teruno Nakaguma[1,3,4], Chitose Ishii[2,4], Arisa Tsuboi[3], Chie Shindo[6], Tamotsu Kato[2], Motoaki Udagawa[9], Takashi Satoh[10], Satoshi Wada[7], Hiroshi Masuya[8], Hisashi Miyamoto[4,11], Hiroshi Ohno[2], Jun Kikuchi*[12]

[1]Graduate School of Horticulture, Chiba University: Matsudo, Chiba 271-8501, Japan

[2]RIKEN Center for Integrated Medical Science : Yokohama, Kanagawa 230-0045, Japan

[3]Japan Eco-science (Nikkan Kagaku) Co. Ltd.: Chiba, Chiba 263-8522, Japan

[4]Sermas Co., Ltd.: Ichikawa, Chiba 272-0033, Japan

[5]Kawachi Suisan Ltd.: Saiki, Oita, 876-2302, Japan

[6]Research Center for Agricultural Information Technology, National Agriculture and Food Research Organization, Tsukuba, Ibaraki 305-0856, Japan,

[7]RIKEN, Center for Advanced Photonics, Wako, Saitama 351-0198, Japan,

[8]RIKEN, BioResource Research Center, Tsukuba, Ibaraki 305-0074, Japan,

[9]Keiyo gas energy solution Co. Ltd.: Ichikawa, Chiba 272-0033, Japan

[10]Division of Hematology, Kitasato University School of Allied Health Sciences, Sagamihara, Kanagawa 252-0329, Japan

[11]Miroku Co.Ltd.: Kitsuki, Oita 873-0021, Japan

[12]RIKEN Center for Sustainable Resource Science: Yokohama, Kanagawa 230-0045, Japan

Corresponding authors. Email: h-miyamoto@faculty.chiba-u.jp; jun.kikuchi@riken.jp

# Figures

**Fig. S1**
The conceptual figure summarizes what was carried out in this study.
**Fig. S2**
Locations where the distribution of seagrass and seaweed was investigated.
**Fig. S3**
The distribution of seagrass and seaweed surveyed in 2004 and 2016 in Oita prefecture is shown in the original public reports.
**Fig. S4**
Photograph of the onshore aquaculture facility in the presence of seagrass Zostera in February 2022.
**Fig. S5**
Photo of the coast around area No. 31 and the shellfish there.
**Fig. S6**
Physicochemical factors of sediments in areas with and without seagrass.
**Fig. S7**
The α-diversity and β-diversity of sediment bacteria in areas with and without seagrass.
**Fig. S8**
Relative abundances of (a) phyla and (b) order of the sediment bacterial population in areas with and without seagrass
**Fig. S9**
Relative abundances of (a) the family and (b) the genera of the sediment bacterial population in areas with and without seagrass
**Fig. S10**
An association network of (a) bacterial orders and (b) a plot of the LDA score and (c) cladogram based on LEfSe (> 1% of each maximum of the detected community) in the area without and with seagrass. (d) Comparison heatmap with bacterial orders selected by AA and LDA.
**Fig. S11**
The seasonal comparison heatmap with bacterial families selected by AA.
**Fig. S12**
Response to Environmental ε based on ELA.
**Fig. S13**
The statistical comparison between clusters of stable states in the entire ELA is linked with (a) the seasons detected in this study and (b) the distance away from the target onshore aquaculture facility.
**Fig. S14**
The bacterial population of the family Rhodobacteraceae in the sediment and the seasonal differences of the genera that belong to them.
**Fig. S15**
The test result for whether the group of bacteria belonging to the optimal structural equation is normally distributed with a QQ plot.
**Fig. S16**
Seasonal causal interaction networks predicted by BayesLiNGAM

# Tables

**Table S1**
The levels of stable isotopes ($\delta^{13}C$ and $\delta^{15}N$), carbon, and nitrogen in the seagrass *Zostera marina* in area No.31

**Table S2**
These changes in the bacterial network would impact the environmental load

**Table S3**
Statistical values of the final optimal structural equation models for bacteria in the sediment the seagrass around the facility administered with the compost

**Table S4**
A list of models targeted by causal mediation analysis for Fig. 2b and their statistical values

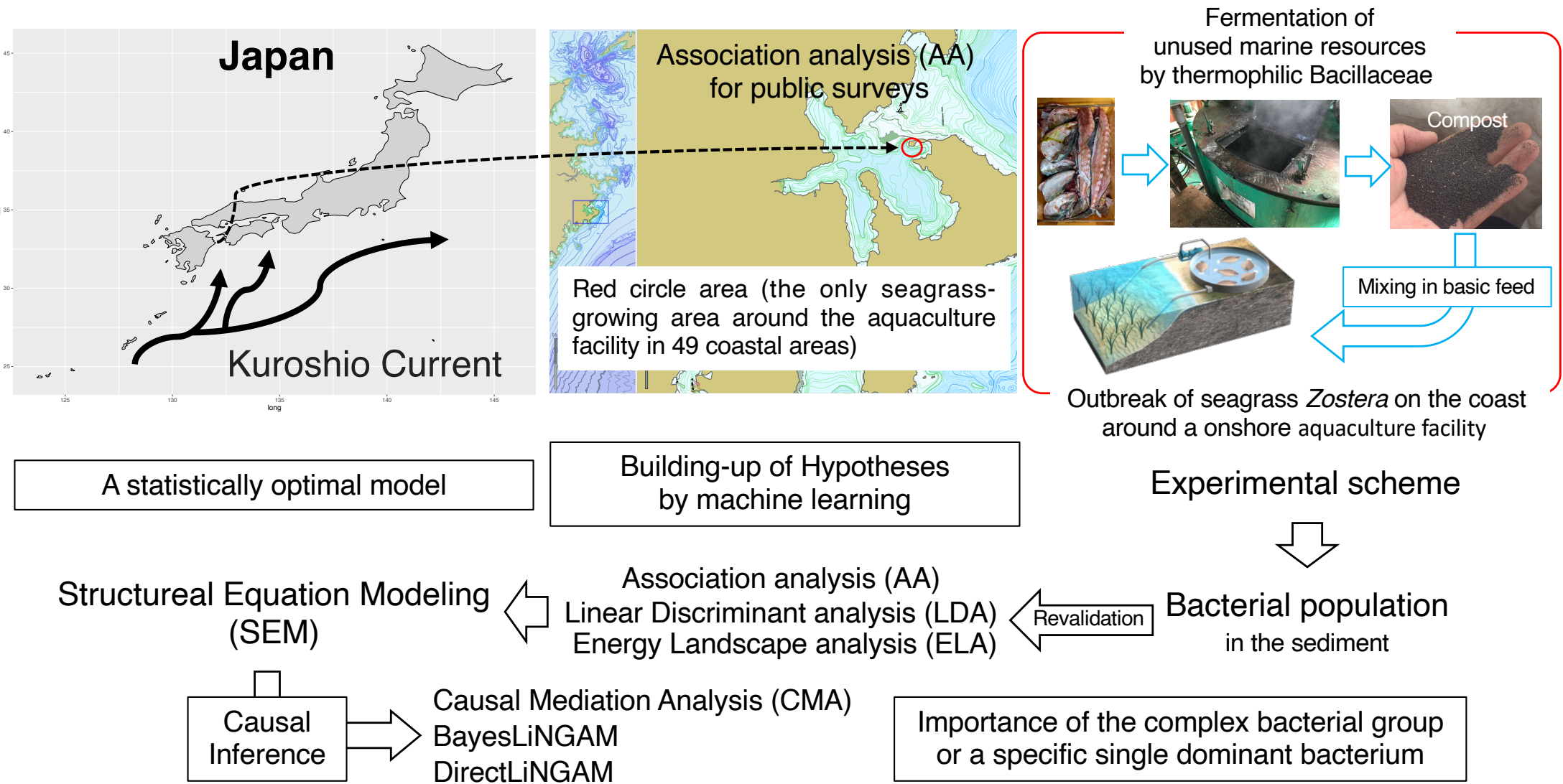

**Fig. S1**
The conceptual figure summarizes what was carried out in this study. As shown in the upper row, public data on seaweed and seagrass distribution along 49 coasts in the aquaculture-rich bay where the Kuroshio tide of Japan enters (surveys in 2004 and 2016) were reanalyzed using association analysis (AA). The red circled area shows the coast around an aquaculture facility. Seagrass was not present in 2004 but was present in 2016. Recycled feed has long been used there. AA, LDA, and ELA investigated the characteristic properties in the sediment. On the basis of these data, the candidates of the bacterial group associated with seagrass growth were estimated by structural equation modeling (SEM) followed by causal mediation analysis (CMA) and a linear non-Gaussian acyclic model (LiNGAM) such as DirectLiNGAM and BayesLiNGAM, methods for statistical causal inference. The importance of the interaction within the bacterial group itself in the group was estimated.

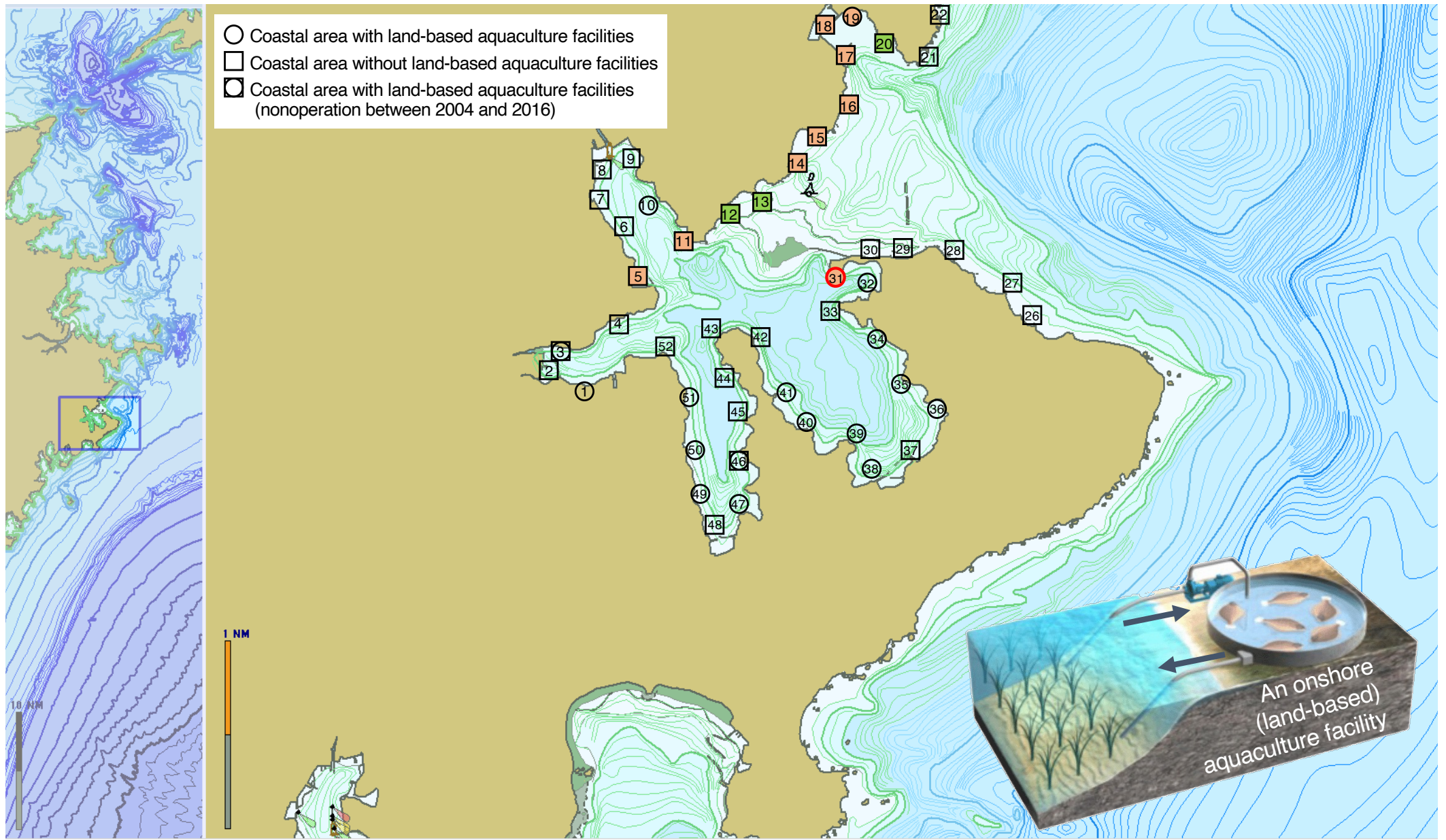

**Fig. S2**
Locations where the distribution of seagrass and seaweed was investigated. The numbers indicate specific locations. Square numbers indicate the absence of nearby onshore aquaculture facilities (conceptual diagram at the top left), and circle numbers indicate the presence of nearby aquaculture facilities. The numbers with squares and circles (areas No. 3 and 46) indicate the facilities that ceased operation between 2004 and 2016. The green color shows the area with seagrass in 2004 and 2016, and the orange color shows the area with seagrass in 2016. The red circle (area No. 31) (the lower right photo) shows aquaculture facilities without seagrass in the neighborhood in 2004. The numbers in this bay are based on reports from public institutions. Area Nos 23-26 are not listed as they are in the margin of the map in this bay.

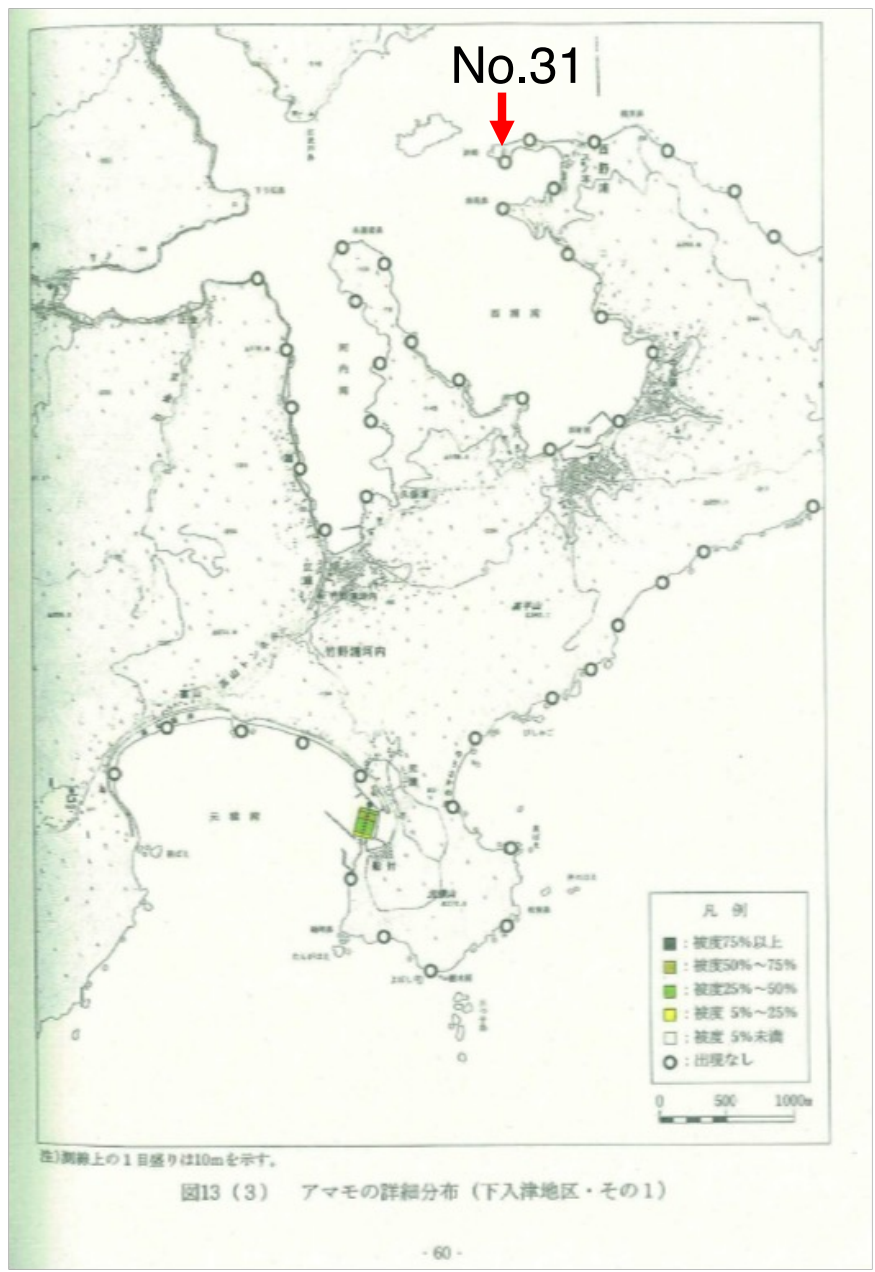
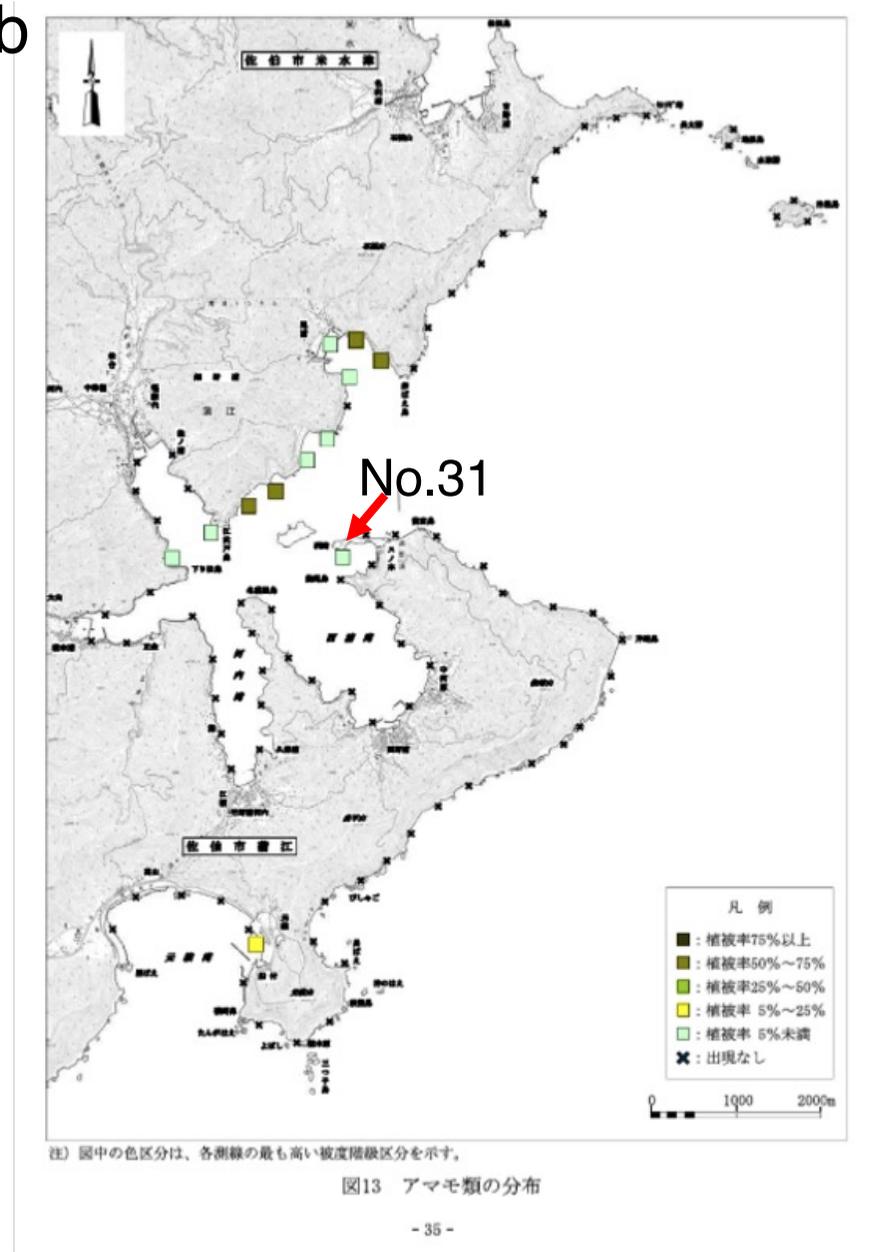

**Fig. S3**
The distribution of seagrass and seaweed surveyed in 2004 and 2016 in Oita prefecture is shown in the original public reports. (a) Page 60 of the original report in 2004 and (b) page 35 of the original report in 2016 were reprinted with permission. Area No. 31 is shown with the red arrow. The circles in (a) show the absence of seagrass, and the squares show the presence of seagrass. The x-mark in (b) shows the absence of seagrass, and the square shows the presence of seagrass.

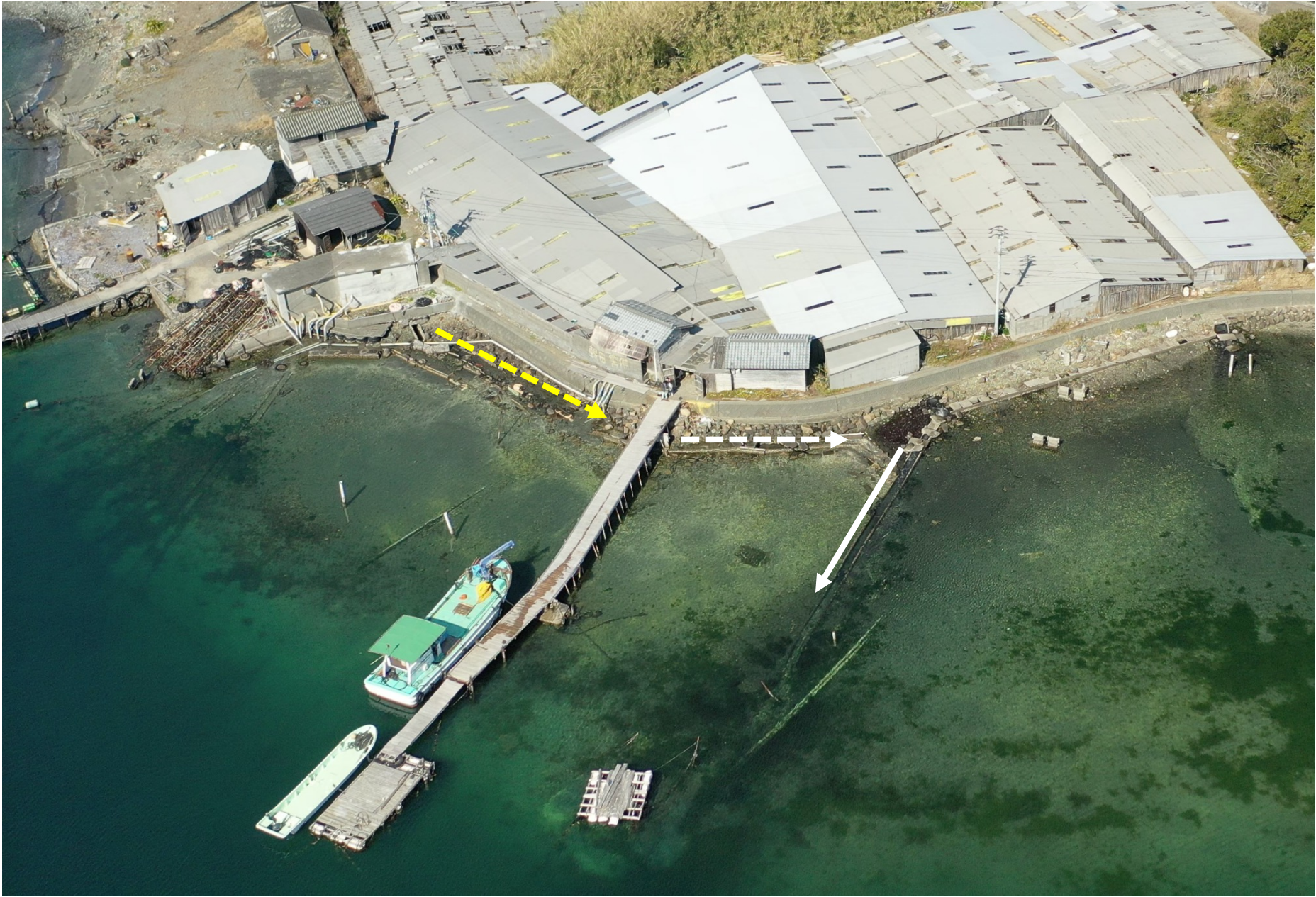

**Fig. S4**
Photograph of the onshore aquaculture facility in the presence of seagrass Zostera in February 2022. The arrow shows the flow of the drainage channel. The dotted arrow indicates the flow of the drainage channel, which is not currently used, in Fig. 1b. The yellow arrow shows the drainage channel in Fig. 1b.

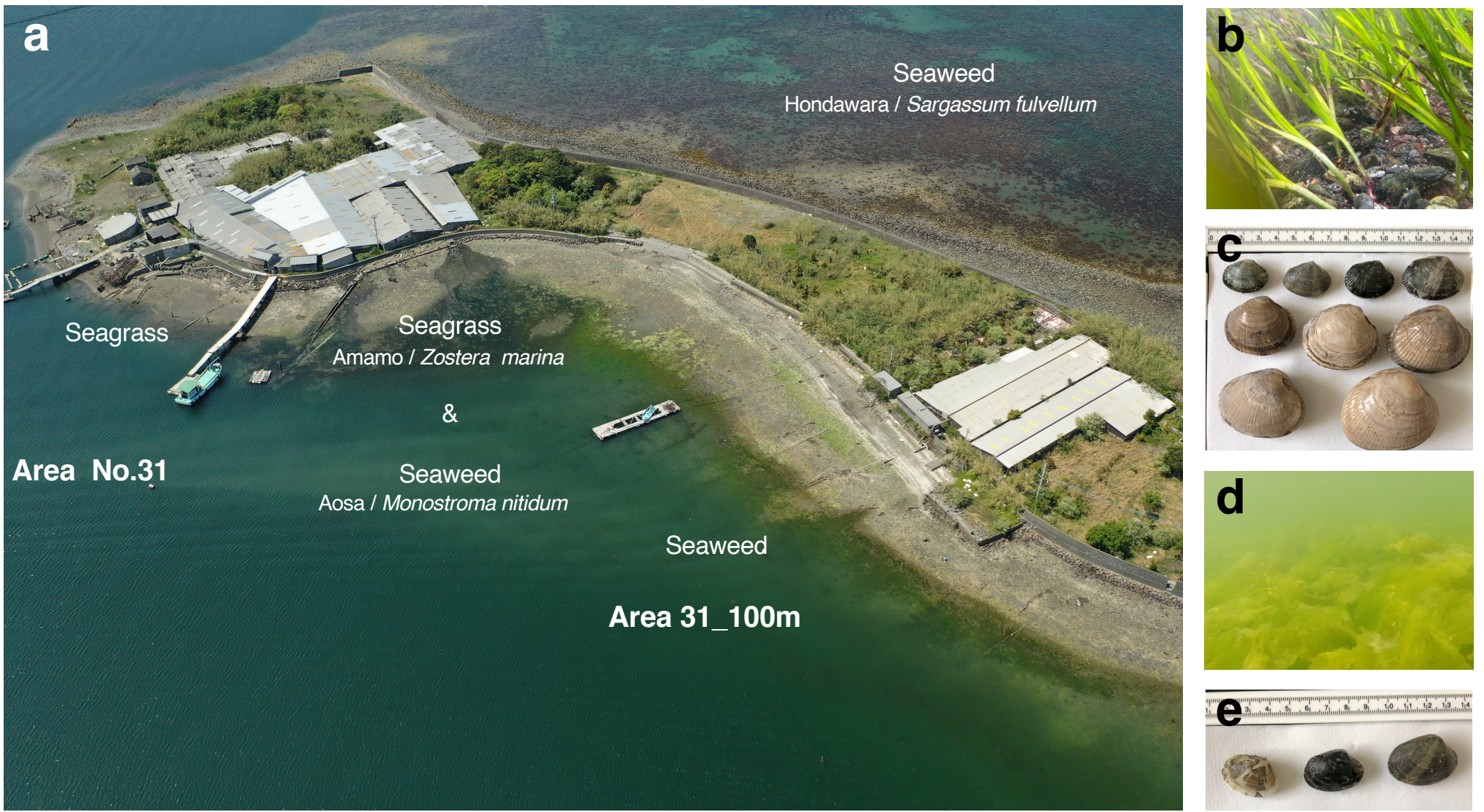

**Fig. S5**
Photo of the coast around area No. 31 and the shellfish there. (a) A photo of the coast around area No. 31 is shown. The photo was taken at spring tide, April 2022. The part where seagrass and/or seaweed grow is marked with words. Photographs of (b) seagrass and (c) shellfish in area No. 31, (d) seaweed and (e) in "area 31_100 m" 100 m from the aquaculture facility close to area No. 31. *Ruditapes philippinarum, Saxidomus purpurata, Fulvia mutica* and *Mya arenaria*, as the appearance of the shellfish was surveyed. (c) shows nine types of shellfish that could be obtained in a radius of approximately 50 cm in 10 minutes. Since the number of shellfish was high, different types of shellfish were gathered. (e) shows three shellfish that could be obtained in a radius of approximately 5 m in 30 minutes around the area without seagrass. A ruler was placed above the photograph (unit: cm).

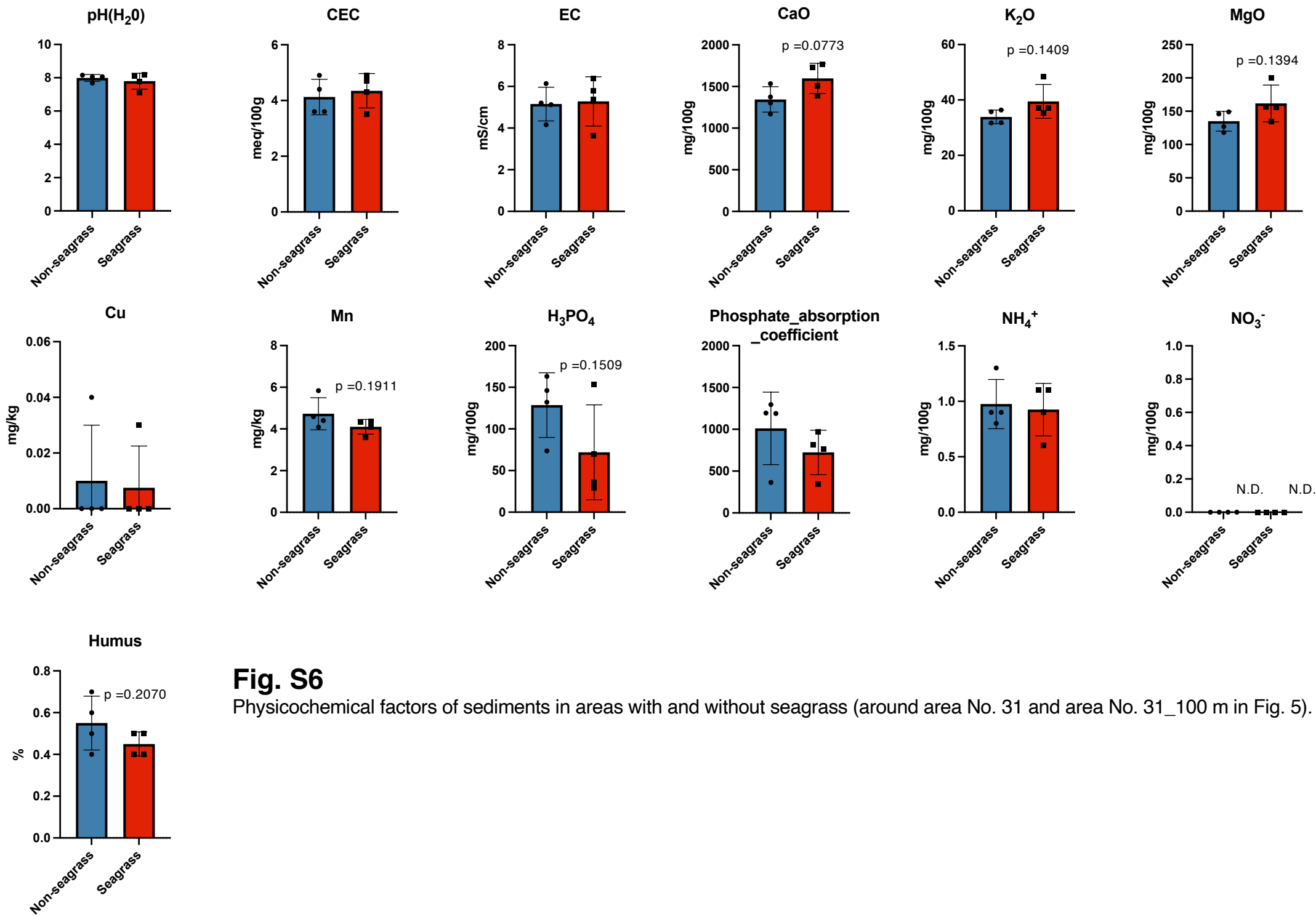

**Fig. S6**
Physicochemical factors of sediments in areas with and without seagrass (around area No. 31 and area No. 31_100 m in Fig. 5).

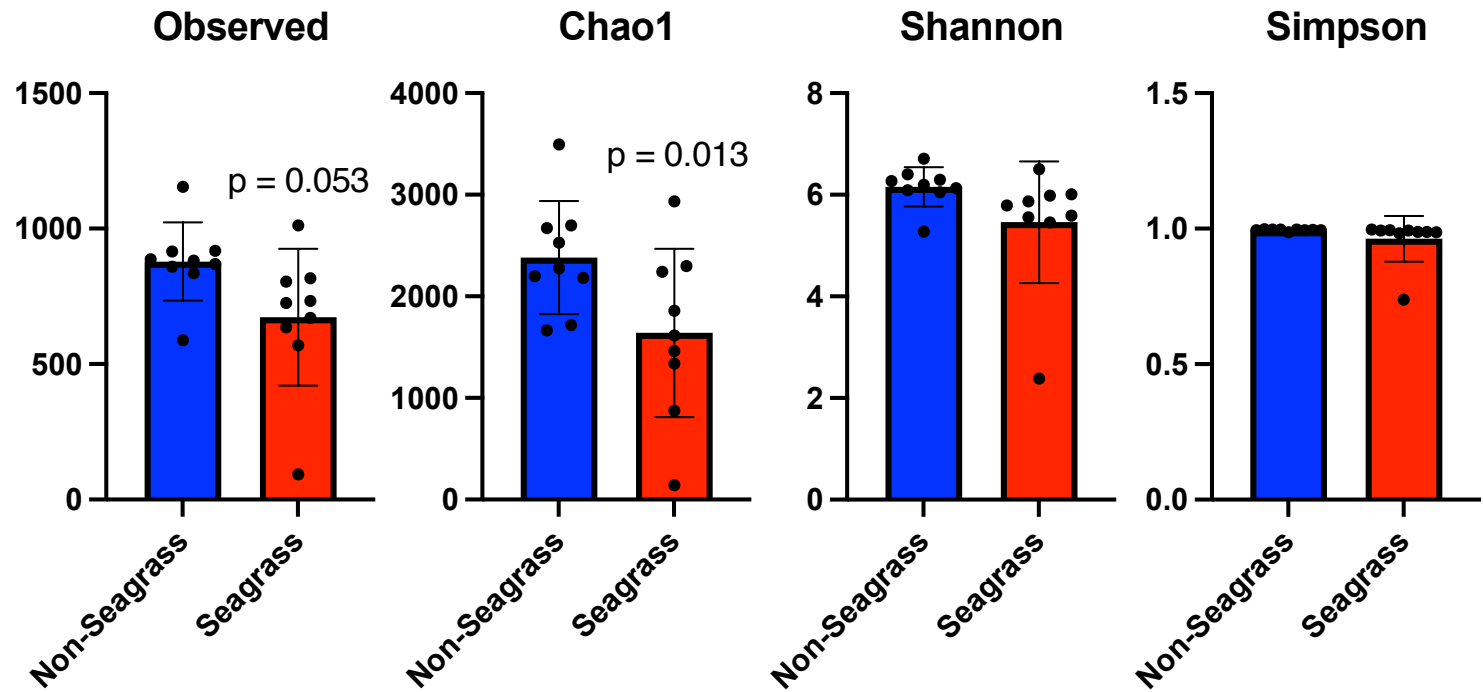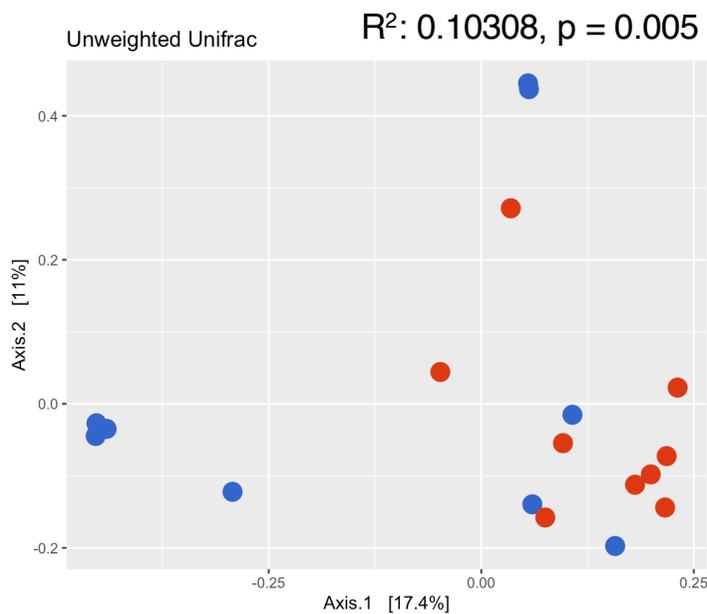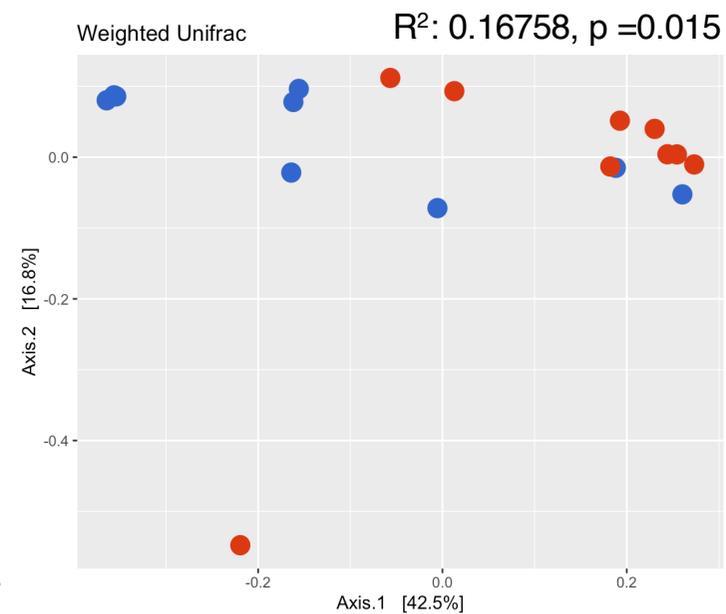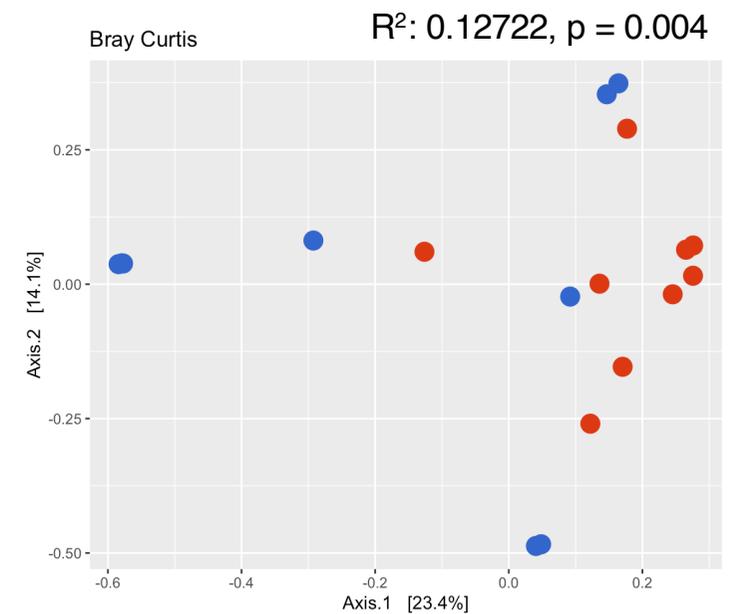

**Fig. S7**

The α-diversity and β-diversity of sediment bacteria in areas with and without seagrass. (a) The number of OTUs observed (observed), Chao1, Simpson, and Shannon indices representing the α-diversity of sediment bacteria in areas with and without seagrass. (b) Unweighted UniFrac graph, (c) weighted UniFrac graph, and (d) Bray–Curtis graph, which shows the β-diversities between the same groups.

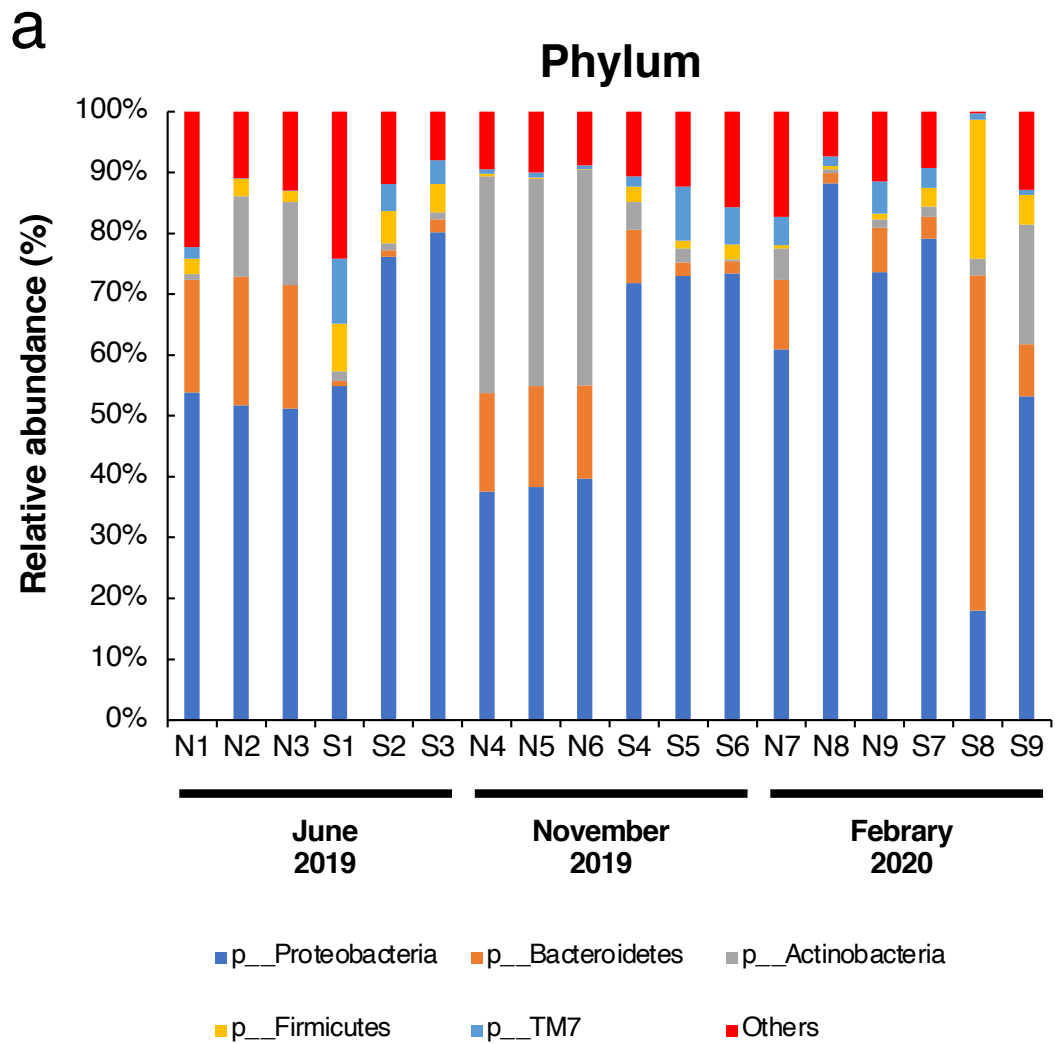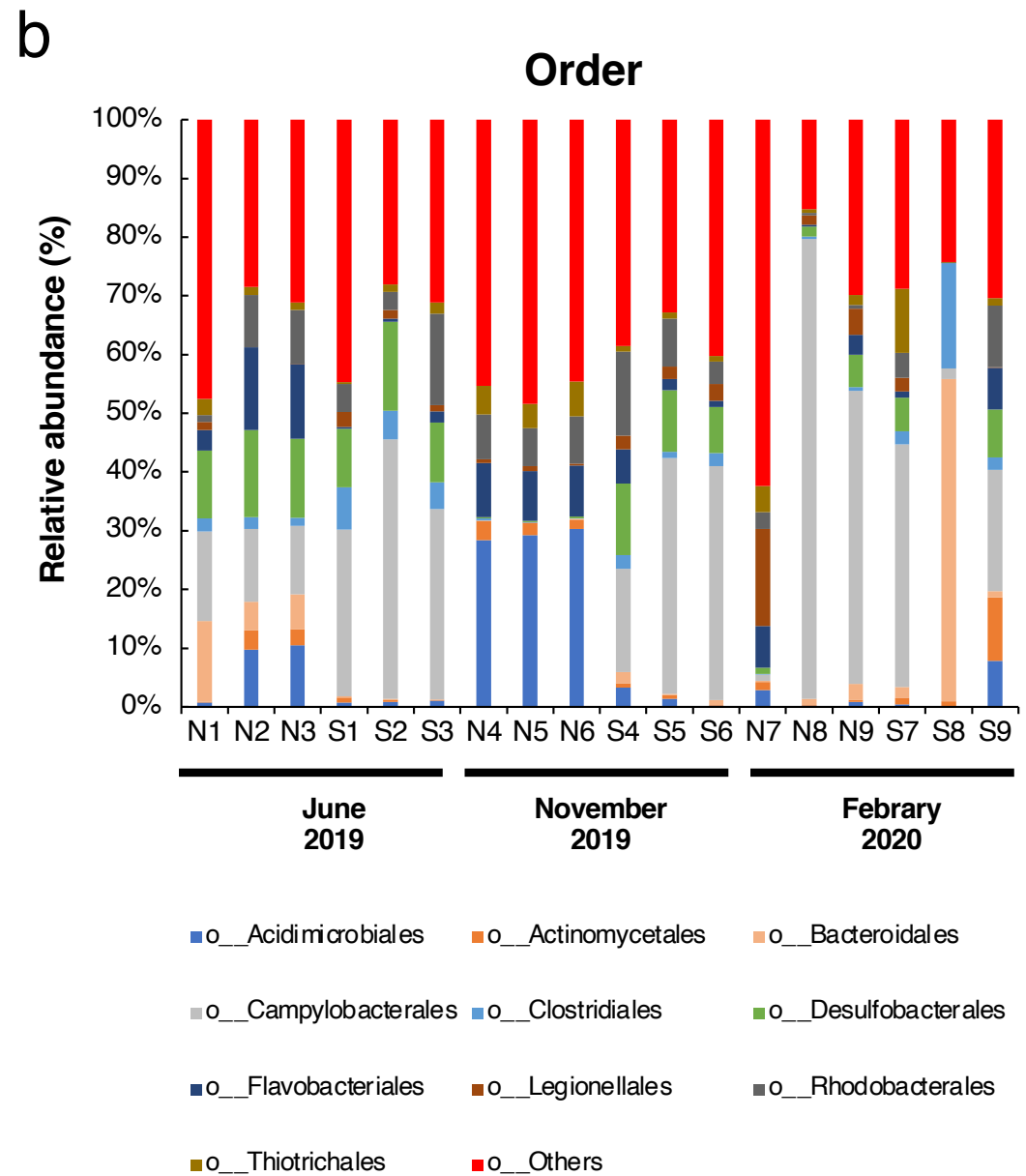

**Fig. S8**
Relative abundances of (a) phyla and (b) order of the sediment bacterial population in areas with and without seagrass (> 10% of the detected community). The abbreviations are as follows: N, Non_seagrass (data in area without seagrass); S, Seagrass (data in area with seagrass); p__, phylum; o__, order. N1-9 and S1-9 show the sample number.

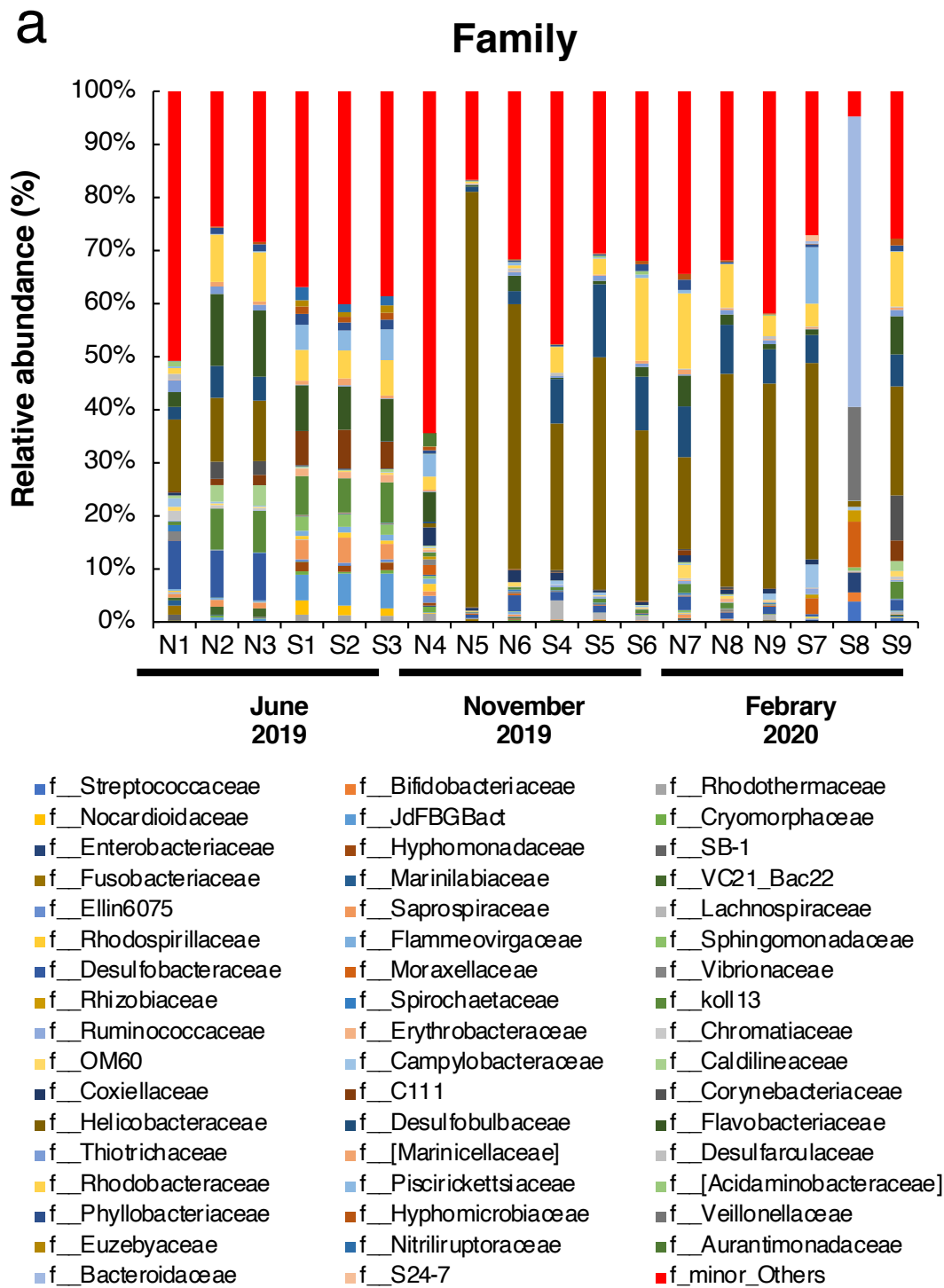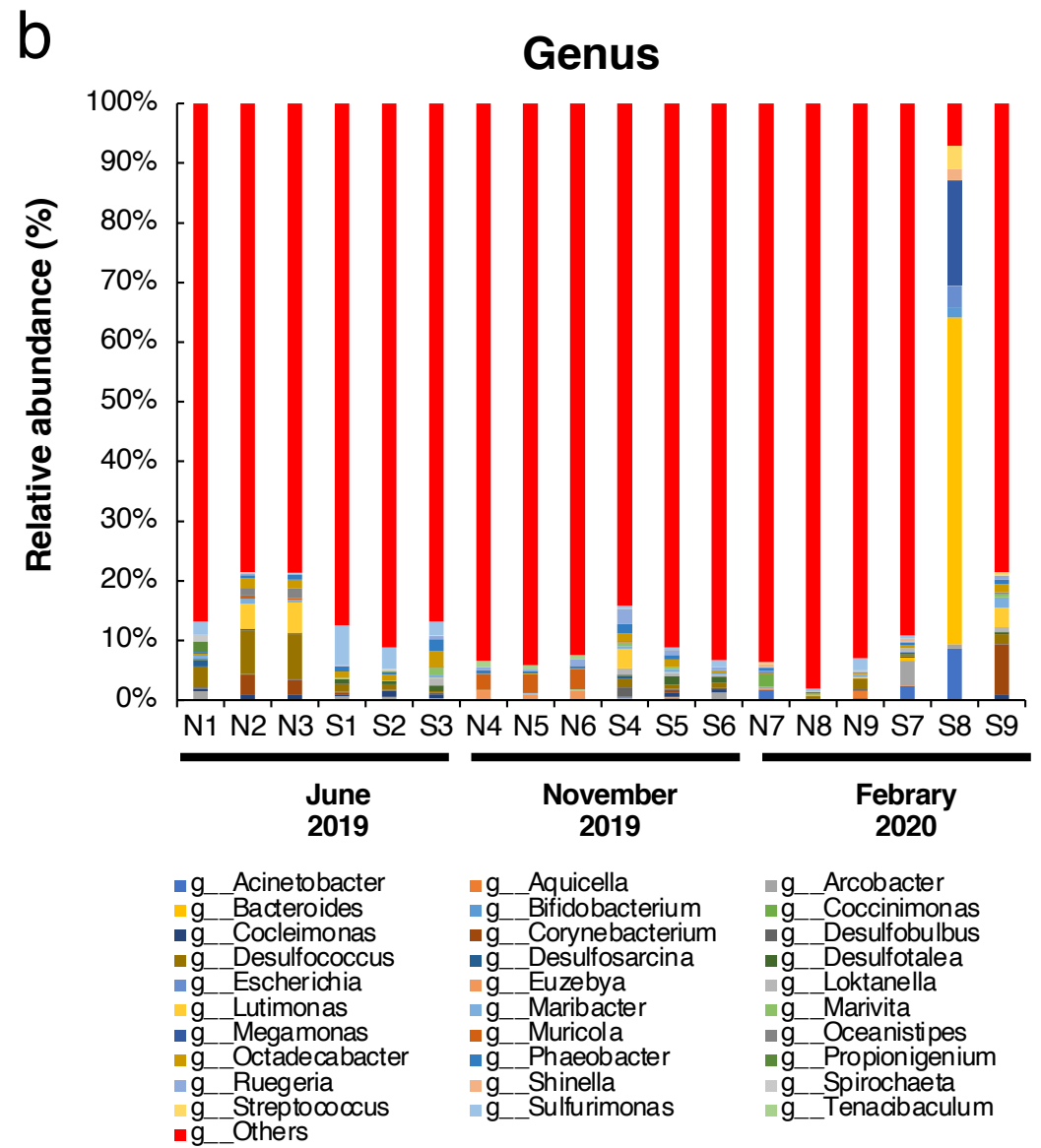

Fig. S9
Relative abundances of (a) the family and (b) the genera of the sediment bacterial population in areas with and without seagrass (> 1% of the detected community). The abbreviations are as follows: N, Non_seagrass (data in area without seagrass); S, Seagrass (data in area with seagrass); f__, family; g__, genus. N1-9 and S1-9 show the sample number.

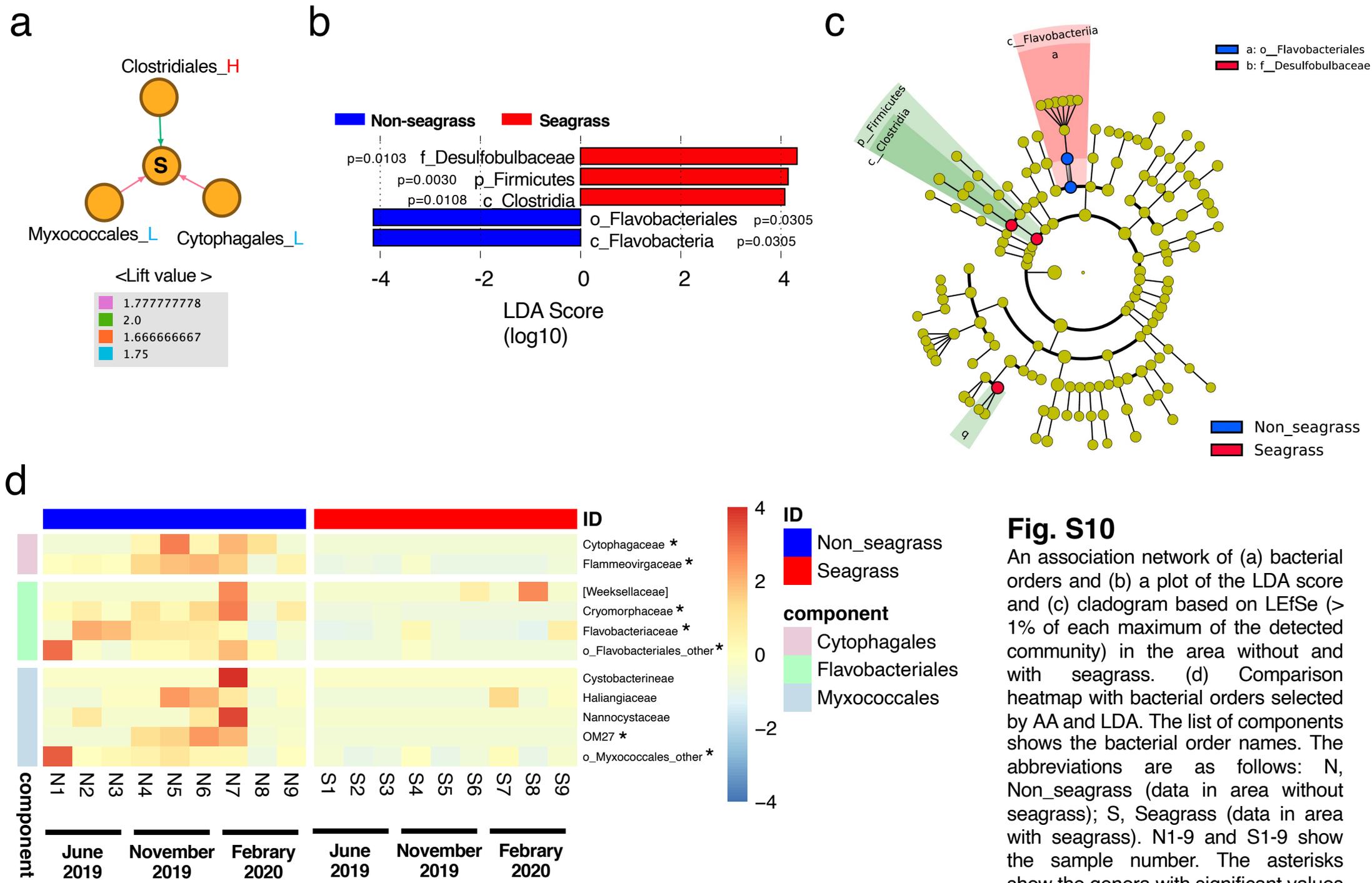

**Fig. S10**
An association network of (a) bacterial orders and (b) a plot of the LDA score and (c) cladogram based on LEfSe (> 1% of each maximum of the detected community) in the area without and with seagrass. (d) Comparison heatmap with bacterial orders selected by AA and LDA. The list of components shows the bacterial order names. The abbreviations are as follows: N, Non_seagrass (data in area without seagrass); S, Seagrass (data in area with seagrass). N1-9 and S1-9 show the sample number. The asterisks show the genera with significant values (p<0.05).

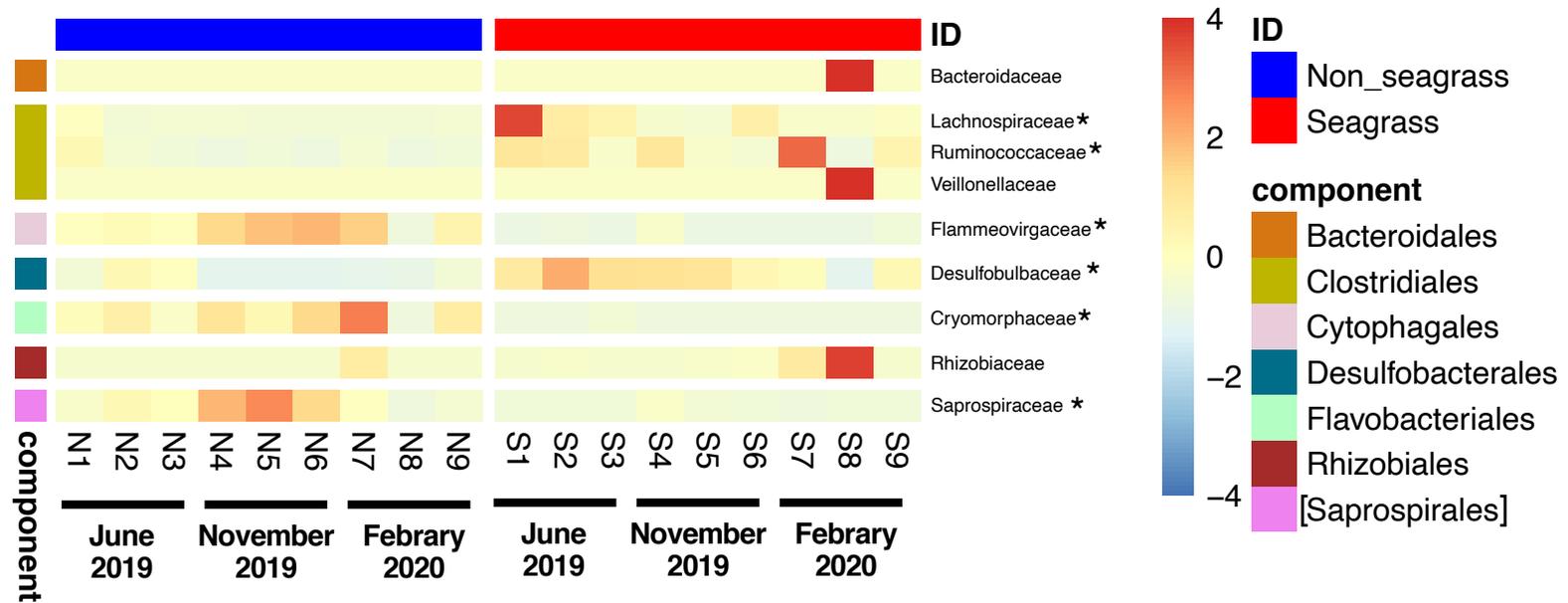

**Fig. S11**
The seasonal comparison heatmap with bacterial families selected by AA. The list of components shows the bacterial order names. The abbreviations are as follows: N, Non_seagrass (data in area without seagrass); S, Seagrass (data in area with seagrass). Asterisks show genera with significant values for all data (p<0.05).

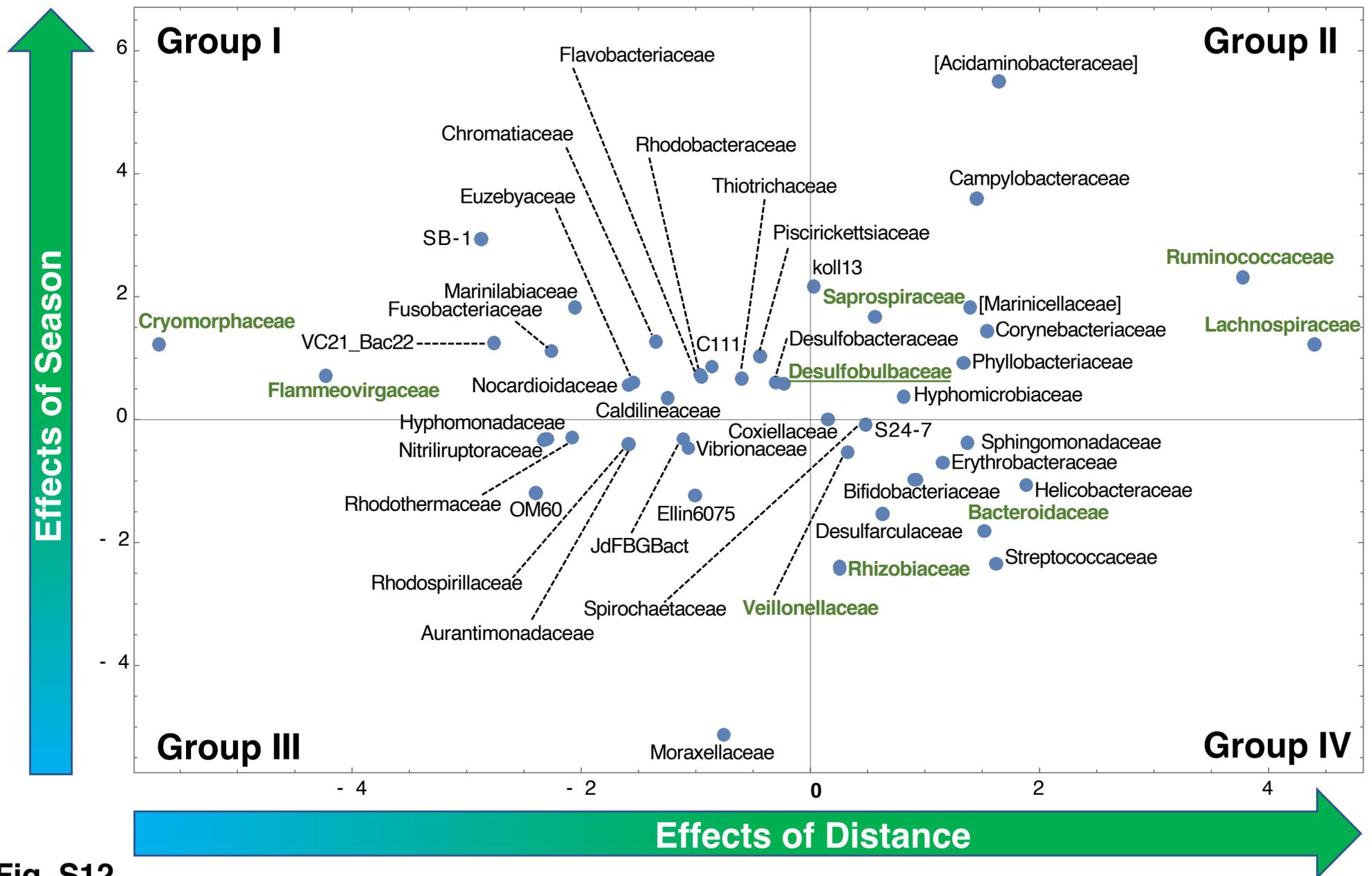

**Fig. S12**
Response to Environmental ε based on ELA. Dependencies on the distance from the target onshore aquaculture facility ($g_i^d$) (X axis) and response to seagrass growth treatment ($g_i^s$) (Y-axis) were plotted. Four categories are shown as Groups I-IV. The upper side of the X-axis shows that it is strongly influenced by the summer (Jun) period, and the lower side is strongly influenced by the winter (Feb) period. Furthermore, the right side of the Y axis shows that it is strongly influenced by the regional character of the coastal area near the aquaculture facility (Area No.31 in Fig. S6), and the left side shows that it is strongly influenced by the regional character of the area No.31_100 m (Fig. S6), which is far from the aquaculture facility. Families with green letters show the components selected by AA. The underlined family shows the component by LDA.

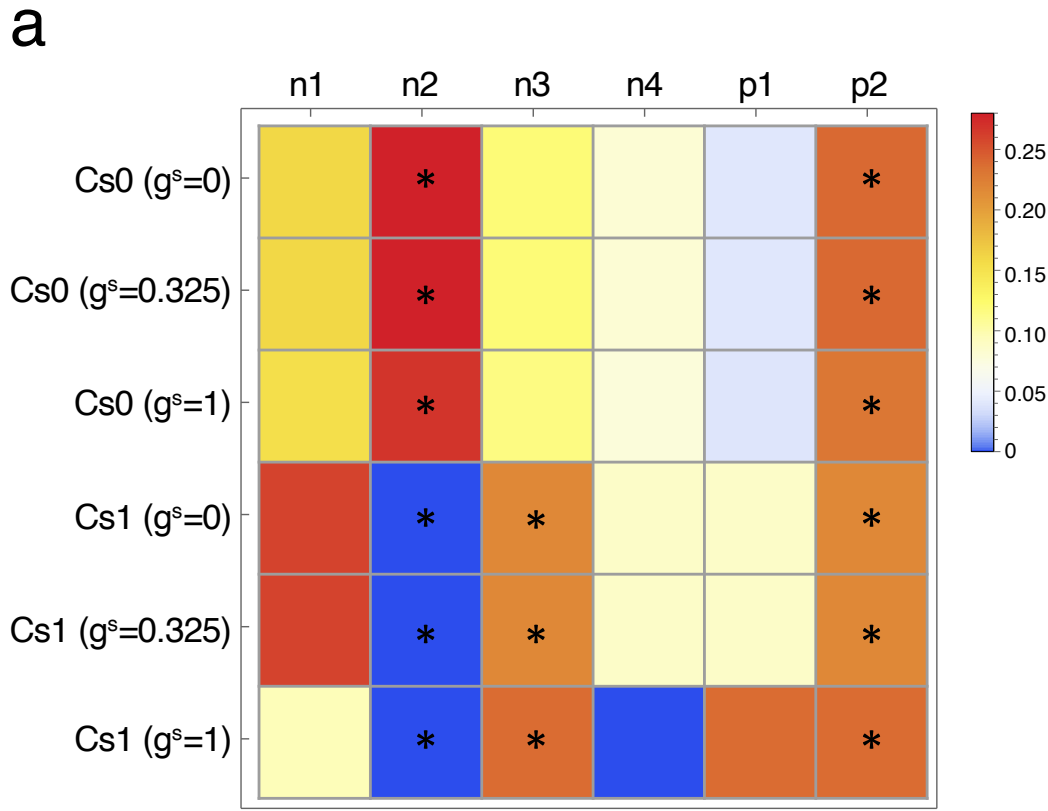 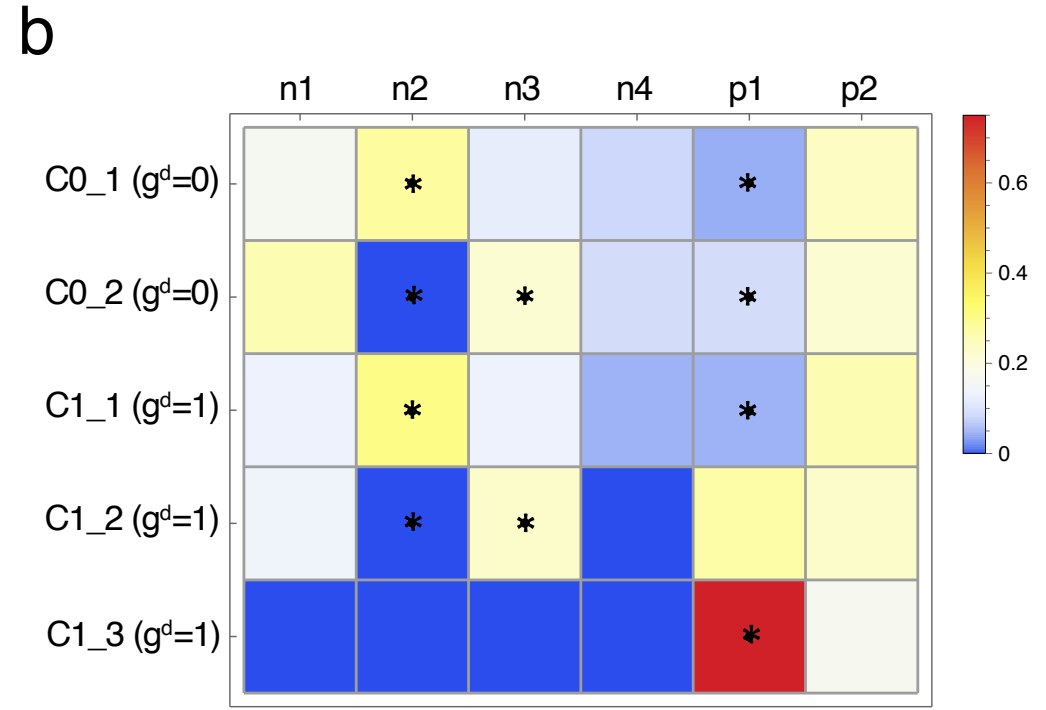

**Fig. S13**
The statistical comparison between clusters of stable states in the entire ELA is linked with (a) the seasons detected in this study and (b) the distance away from the target onshore aquaculture facility. Labels of n1, n2, n3, n4, p1, and p2 show the names of bacterial clusters calculated by ELA. (a) The seasonal values are as follows: $g^s=0$, February; $g^s=0.325$, November; and $g^s=1$, June. Cs0 and Cs1 show the labels of seasonal stable states shown in Fig. 4a. The labels of Cs0($g^s=0$), Cs0($g^s=0.325$), and Cs0($g^s=1$) show the names of the season-based clusters in ELA. Cs0 ($g^s = 0$), Cs0 ($g^s = 0.325$), and Cs0($g^s=1$) show stable Cs0 clusters in February, November, and June. (b) The labels of the y axis (C0_1, C0_2, C1_1, C1_2, and C1_3) show the stable states in Fig. 4a. The distance values are as follows: $g^d=0$, area No.31_100 m; and $g^d=1$, area No.31 in Fig. S6, respectively. The asterisks show the statistical significance ($p<0.05$) for the stable states of the calculated energy landscape (Fig. 4).

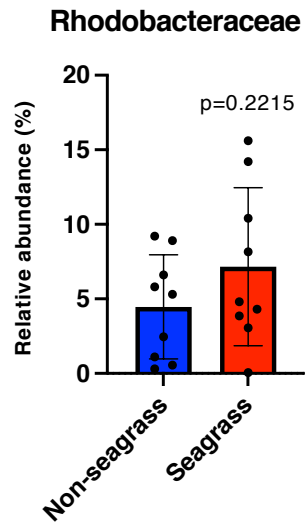

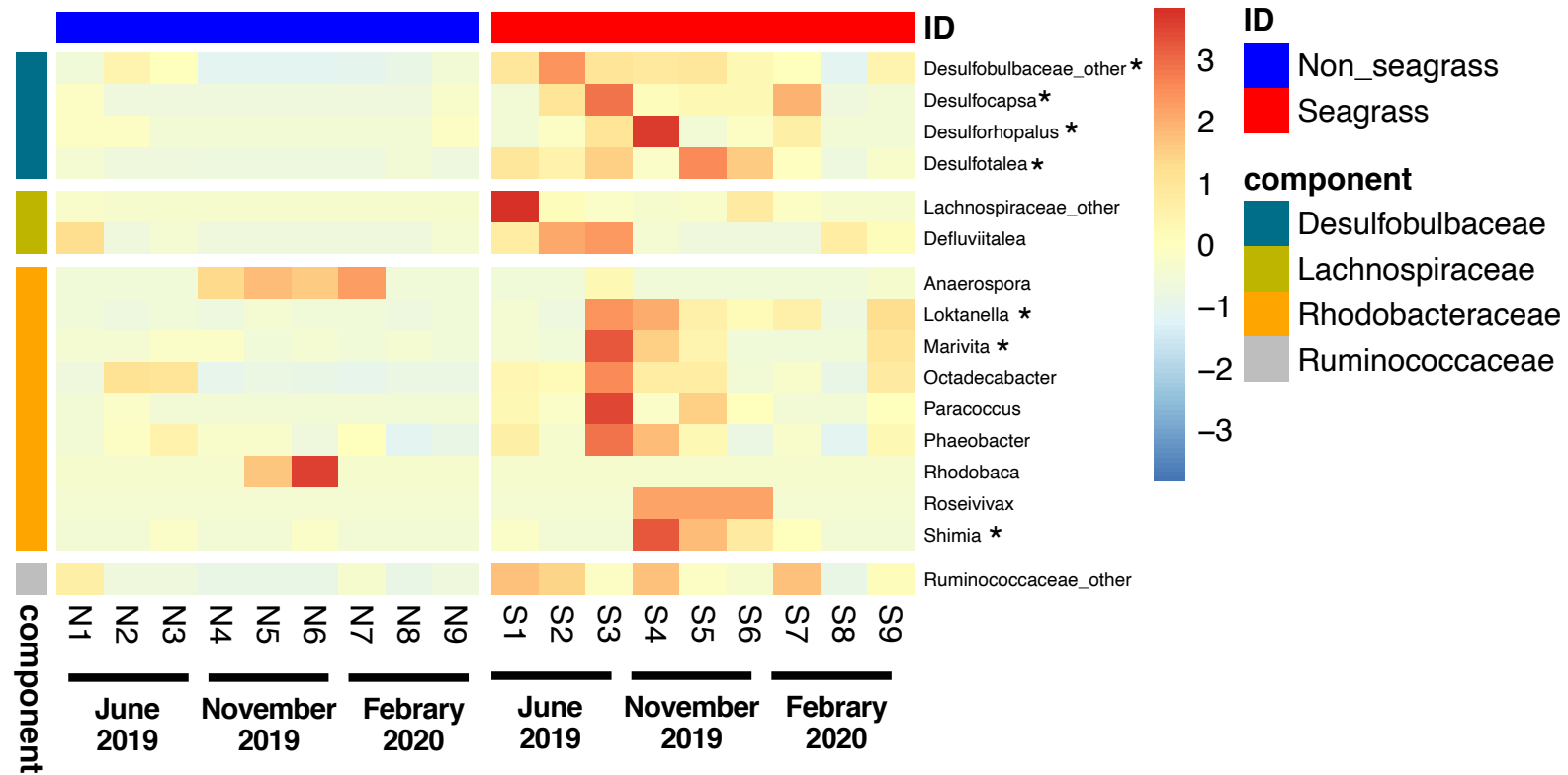

**Fig. S14**
The bacterial population of the family Rhodobacteraceae in the sediment and the seasonal differences of the genera that belong to them. (a) Relative abundance of the family Rhodobacteraceae. (b) Seasonal comparison heatmap representing the predominantly changed bacterial genera belonging to the families. The list of components shows the bacterial family names. The abbreviations are as follows: N, Non_seagrass (data in area without seagrass); S, Seagrass (data in area with seagrass). Asterisks show genera with significant values ($p<0.05$).

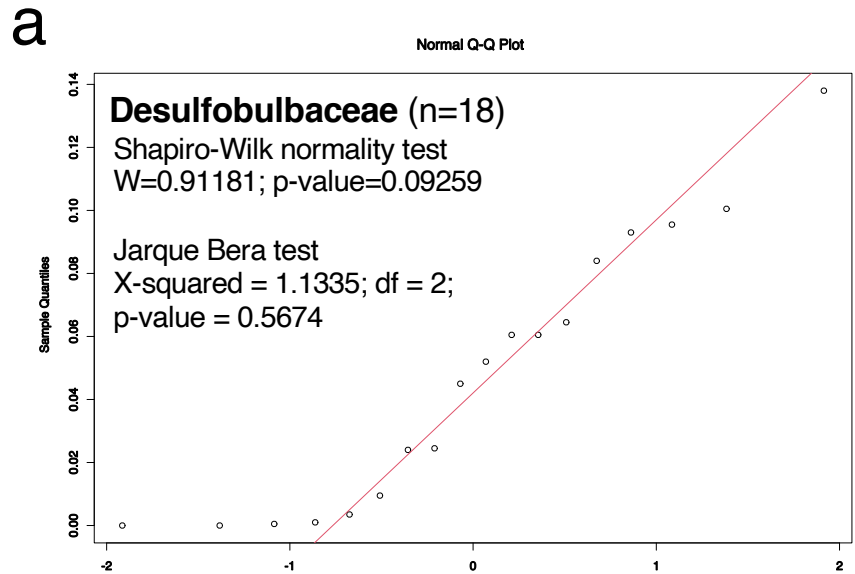
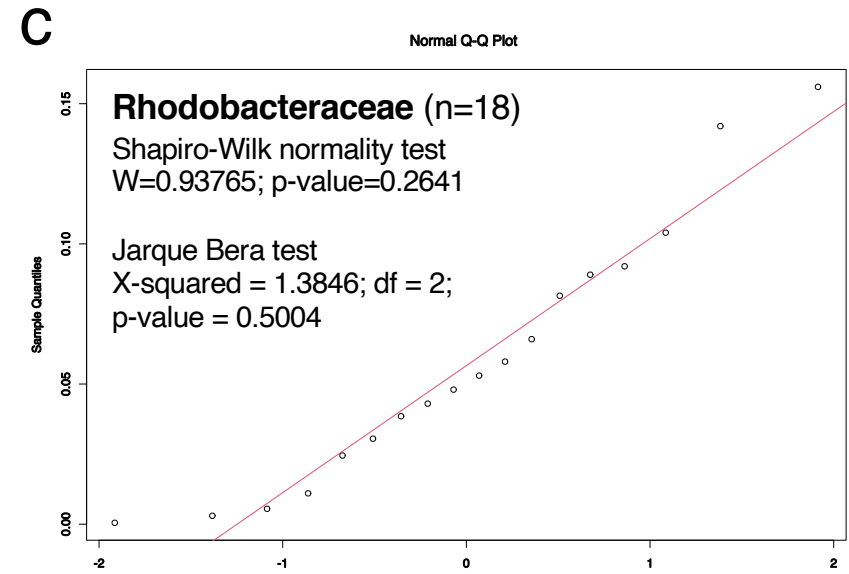
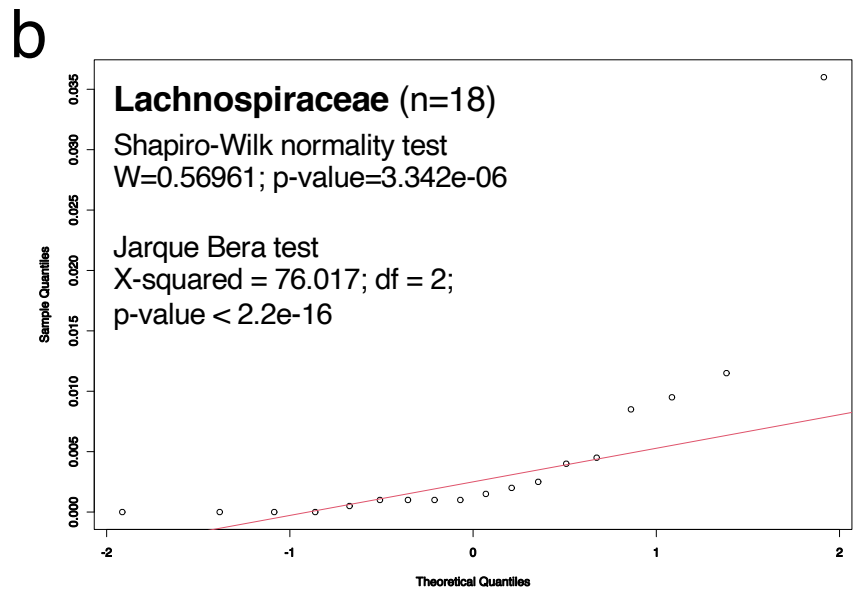
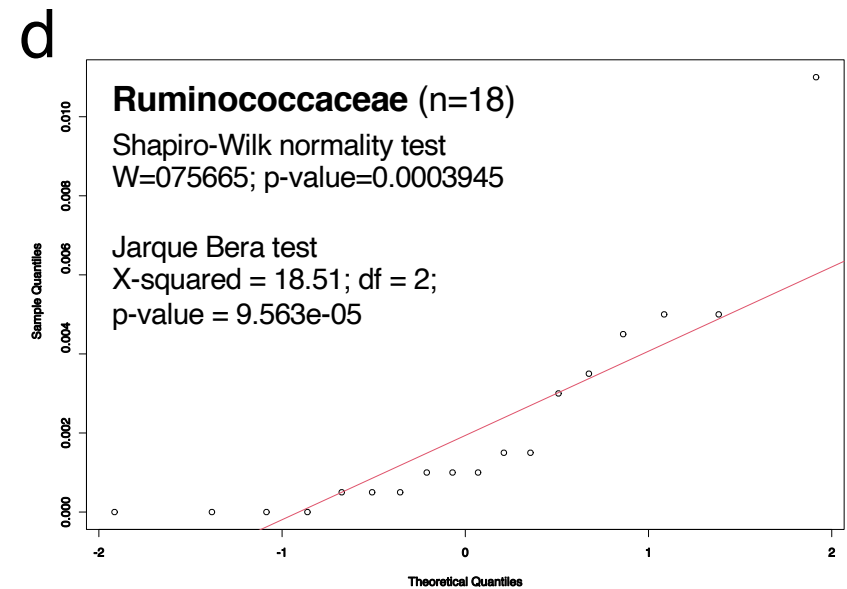

**Fig. S15**
The test result for whether the group of bacteria belonging to the optimal structural equation is normally distributed with each QQ plot: (a)Desulfobulbaceae; (b) Lachnospiraceae; (c) Rhodobacteraceae; and (d) Ruminococcaceae. Raw data for the bacterial families were calculated using the Shapiro-Wilk normality test and the Jarque Bera test (n=18).

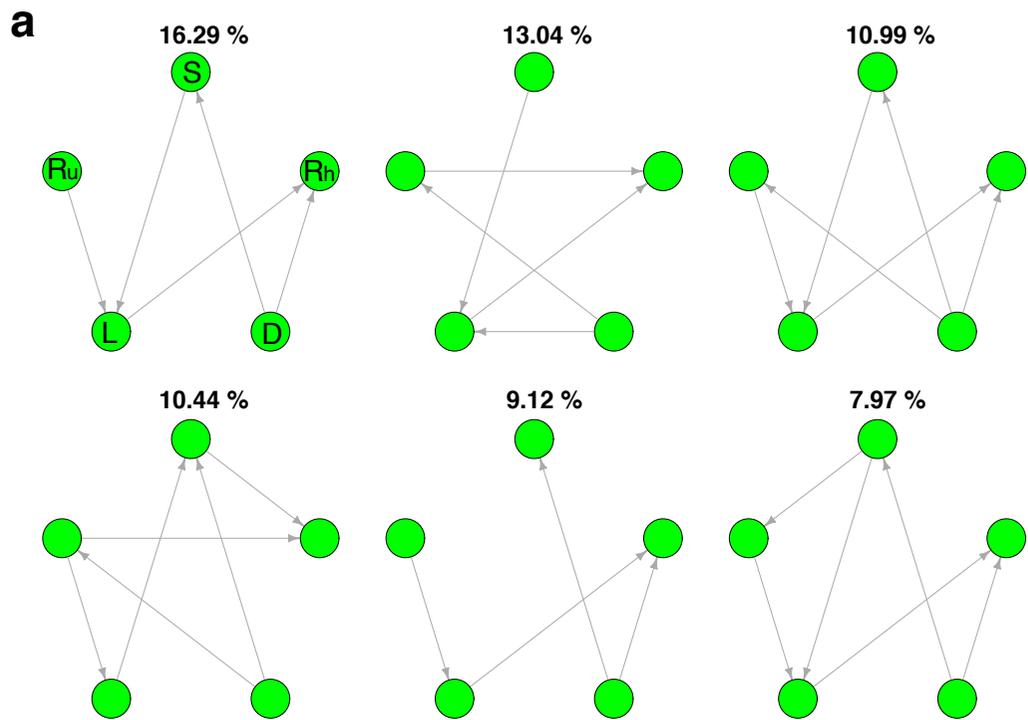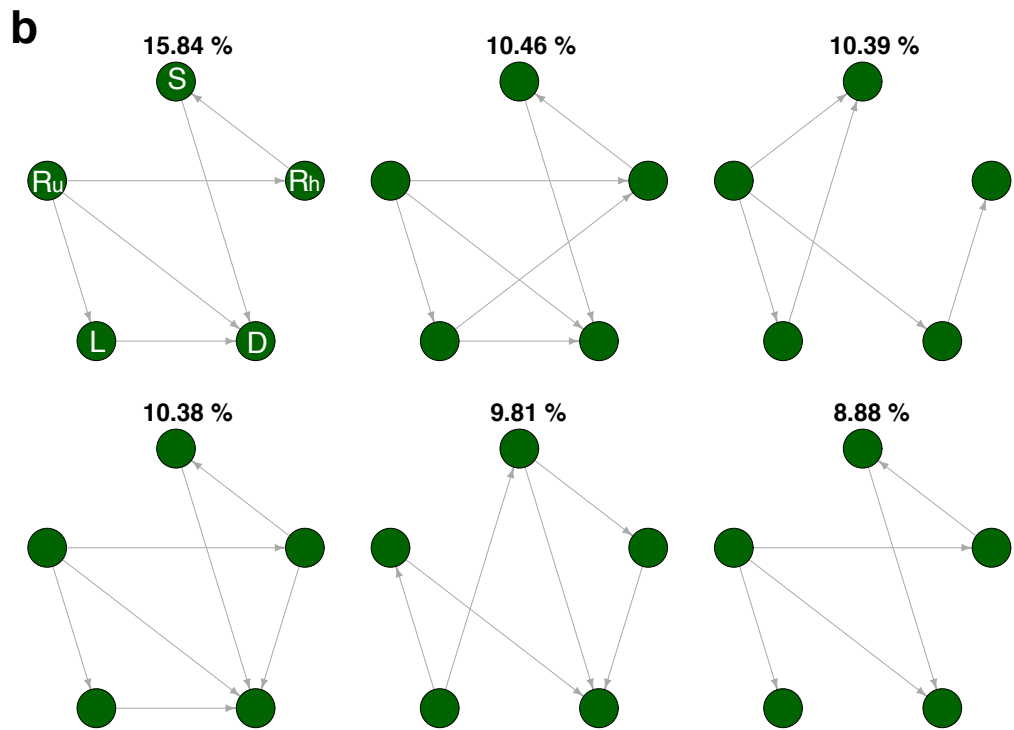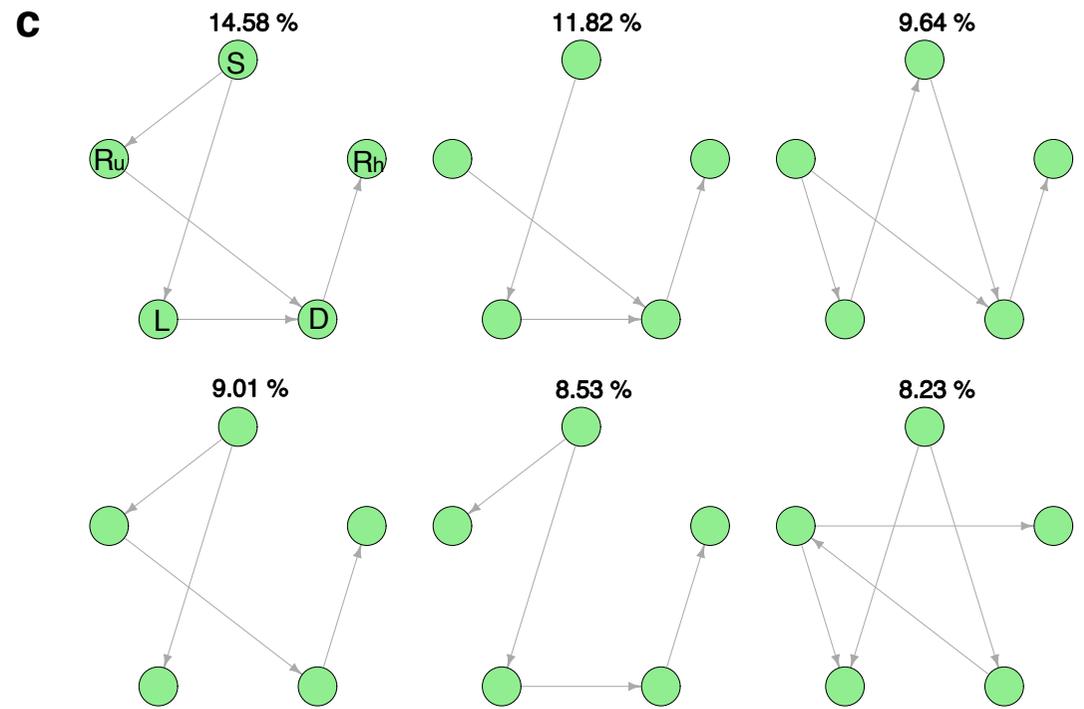

**Fig. S16**

Seasonal causal interaction networks predicted by BayesLiNGAM. The calculated networks are shown in (a) June, (b) November, and (c) February. Abbreviations in the table indicate the following: S, Seagrass; Ru, family Ruminococcaceae; L, family Lachnospiraceae; D, family Desulfobulbaceae; Rh, Rhodobacteraceae.

**Table S1**

The levels of stable isotopes ($\delta^{13}C$ and $\delta^{15}N$), carbon, and nitrogen of the seagrass *Zostera marina* in area No.31.

| Sample No | $\delta^{13}C$ (‰) | $\delta^{15}N$ (‰) | Total N (%) | Total C (%) |
|---|---|---|---|---|
| 1 | -12.2 | 7.3 | 2.2 | 30.2 |
| 2 | -11.4 | 7.3 | 1.7 | 27.2 |
| 3 | -12.0 | 6.7 | 1.8 | 26.1 |
| Av. | -11.9 ± 0.2 | 7.1 ± 0.2 | 1.9 ± 0.2 | 27.8 ± 1.2 |

**Table S2**

These changes in the bacterial network would impact the environmental load. A water quality analysis showed almost no differences in water quality between filtered groundwater and circulating seawater near the onshore aquaculture facility.

| Condition | COD (mg/L) | Total N (mg/L) | Total P (mg/L) | $NH_4^+$ (mg/L) | $NO_3^-$ (mg/L) | SS (mg/L) |
|---|---|---|---|---|---|---|
| filtered sea water | 0.5 | 0.44 | 0.046 | 0.08 | 0.54 | <1 |
| non-filtered sea water | 0.7 | 0.48 | 0.068 | 0.14 | 0.1 | <1 |

# Table S3

Statistical values of the final optimal structural equation models for bacteria in the sediment of seagrass around the target aquaculture facility. The bacteria as the SEM component were selected from AA, LDA, and ELA. Column no. 1 shows the best numerical SEM in Fig. 5a. Columns 2-6 show the models of inferior numerical structural equations. The abbreviations in the table mean the following: chisq, chi-square $\chi^2$; df, degrees of freedom; p value, p values (chi-square); cfi, comparative fix index (CFI); tli, Tucker–Lewis index (TLI); nfi, normed fit index (NFI); rfi, relative fit index (RFI); srmr, standardized root mean residuals (SRMR); AIC, Akaike information criterion; rmsea, root mean square error of approximation (RMSEA); gfi, goodness-of-fit index (GFI); agfi, adjusted goodness-of-fit index (AGFI). The underlines were shown as values inferior to those of other models with goodness-of-fit indices.

| Category | Model | Fit indices | | |
|---|---|---|---|---|
| No.1 | Seagrass ~ Ruminococcaceae + Lachnospiraceae + Desulfobulbaceae<br>Lachnospiraceae ~ Rhodobacteraceae + Desulfobulbaceae<br>lavaan 0.6-11 ended normally after 2 interactions<br>Number of successful bootstrap draws: 828 | chisq 0.149<br>cfi 1.000<br>nfi 0.993<br>rmsea 0.000 | df 2.000<br>tli 1.457<br>srmr 0.017<br>gfi 0.996 | pvalue 0.928<br>rfi 0.975<br>AIC -102.203<br>agfi 0.967 |
| No.2 | Seagrass ~ Ruminococcaceae + Lachnospiraceae + Desulfobulbaceae +<br>Bacteroidaceae + Veillonellaceae<br>Lachnospiraceae ~ Rhodobacteraceae + Desulfobulbaceae<br>lavaan 0.6-11 ended normally after 1 interactions<br><u>Number of successful bootstrap draws: 478</u> | chisq 1.031<br>cfi 1.000<br>nfi 0.978<br>rmsea 0.000 | df 4.000<br>tli 1.226<br>srmr 0.015<br>gfi 0.993 | pvalue 0.905<br><u>rfi 0.940</u><br>AIC -113.032<br>agfi 0.951 |
| No.3 | Seagrass + Rhodobacterales ~ Flavobacteriales<br>Clostridiales + Rhodobacterales ~ Seagrass<br>Clostridiales ~ Rhodobacterales<br>Seagrass ~ Myxococcales<br>lavaan 0.6-11 ended normally after 1 interactions<br>Number of successful bootstrap draws: 998 | chisq 1.076<br>cfi 1.000<br>nfi 0.973<br>rmsea 0.000 | df 3.000<br>tli 1.185<br>srmr 0.029<br>gfi 0.990 | pvalue 0.783<br><u>rfi 0.920</u><br>AIC -119.913<br>agfi 0.951 |
| No.4 | Flammeovirgaceae ~ Saprospiraceae + Cryomorphaceae<br>Lachnospiraceae + Saprospiraceae ~ seagrass + Desulfobulbaceae<br>lavaan 0.6-11 ended normally after 1 interactions<br><u>Number of successful bootstrap draws: 39</u> | chisq 1.085<br>cfi 1.000<br>nfi 0.986<br>rmsea 0.000 | df 6.000<br>tli 1.152<br>srmr 0.025<br>gfi 0.999 | pvalue 0.982<br>rfi 0.972<br>AIC -431.492<br>agfi 0.997 |
| No.5 | Seagrass ~ Clostridiales + Desulfobulbaceae<br>Clostridiales ~ Paracoccus + Phaeobacter + Desulfobulbaceae<br>lavaan 0.6-11 ended normally after 1 interactions<br>Number of successful bootstrap draws: 952 | chisq 0.459<br>cfi 1.000<br>nfi 0.979<br>rmsea 0.000 | df 2.000<br>tli 1.352<br>srmr 0.019<br>gfi 0.988 | pvalue 0.795<br><u>rfi 0.928</u><br>AIC -45.766<br><u>agfi 0.908</u> |

## Table S4

A list of models targeted by causal mediation analysis for Fig. 5a and their statistical values. The coefficients and the calculated data related to the regression model are shown on the left. The values mediated among the regression models are shown on the right. The number in 'Stimulations' shows the bootstrapping frequency. *, p<0.05; **, p<0.01; #, p <0.1.

| Regression models | | | | | (I) Seagrass ~ Ruminococcaceae + Lachnospiraceae + Desulfobulbaceae<br>(II) Lachnospiraceae ~ Rhodobacteraceae + Desulfobulbaceae | | | | |
|---|---|---|---|---|---|---|---|---|---|
| | | | | | Nonparametric bootstrap Confidence Intervals (T: Desulfobulbaceae-M: Lachnospiraceae) | | | | |
| (I) | Estimate std. | Error | t value | Pr ( > ltl) | (I) - (II) | Estimate 95% | CI Lower 95% | CI Upper | p-value |
| (Intercept) | 0.06096 | 0.13749 | 0.443 | 0.6643 | ACME | 0.000 | 0.0000 | 0.00 | 1.000 |
| Ruminococcaceae | 46.5468 | 36.94729 | 1.26 | 0.2283 | ADE | 0.125 | 0.0325 | 0.34 | 0.004 ** |
| Lachnospiraceae | 8.57616 | 12.03459 | 0.713 | 0.4878 | | | | | |
| Desulfobulbaceae | 6.23402 | 2.60537 | 2.393 | 0.0313 * | Nonparametric bootstrap Confidence Intervals (T: Desulfobulbaceae-M: Rhodobacteraceae) | | | | |
| | | | | | (I) - (II) | Estimate 95% | CI Lower 95% | CI Upper | p-value |
| Residual standard error: 0.3751 on 14 degrees of freedom | | | | | ACME | -0.385 | -1.539 | 0.13 | 0.178 |
| Multiple R-squared: 0.5622, Adjusted R-squared: 0.4683 | | | | | ADE | 0.125 | 0.027 | 0.33 | 0.004 ** |
| F-statistic: 5.992 on 3 and 14 DF, p-value: 0.007606 | | | | | | | | | |
| | | | | | Nonparametric bootstrap Confidence Intervals (T: Desulfobulbaceae-M: Lachnospiraceae) | | | | |
| (II) | Estimate std. | Error | t value | Pr ( > ltl) | (II) - (I) | Estimate 95% | CI Lower 95% | CI Upper | p-value |
| | | | | | ACME | 1.072 | -0.5628 | 9.28 | 0.124 |
| (Intercept) | 0.002339 | 0.003162 | 0.74 | 0.4709 | ADE | 6.234 | -1.7494 | 10.96 | 0.092 # |
| Rhodobacteraceae | -0.061827 | 0.047101 | -1.313 | 0.209 | | | | | |
| Desulfobulbaceae | 0.125003 | 0.050585 | 2.471 | 0.0259 * | Quasi-Bayesian Confidence Intervals (T: Desulfobulbaceae-M: Rhodobacteraceae) | | | | |
| | | | | | (I) - (II) | Estimate 95% | CI Lower 95% | CI Upper | p-value |
| Residual standard error: 0.007685 on 15 degrees of freedom | | | | | ACME | -0.389 | -1.308 | 0.19 | 0.220 |
| Multiple R-squared: 0.2896, Adjusted R-squared: 0.1949 | | | | | ADE | 0.126 | 0.028 | 0.23 | 0.006 ** |
| F-statistic: 3.058 on 2 and 15 DF, p-value: 0.07695 | | | | | | | | | |
| | | | | | Quasi-Bayesian Confidence Intervals (T: Desulfobulbaceae-M: Rhodobacteraceae) | | | | |
| | | | | | (II) - (I) | Estimate 95% | CI Lower 95% | CI Upper | p-value |
| | | | | | ACME | 1.072 | -0.715 | 9.36 | 0.098 # |
| | | | | | ADE | 6.234 | -2.346 | 10.99 | 0.118 |